\documentclass[journal]{IEEEtran}
\usepackage{cite}
\usepackage{amsmath,amssymb,amsfonts}
\usepackage{algorithmic}
\usepackage{graphicx}
\usepackage{textcomp}
\usepackage{stfloats}
\usepackage{makecell}
\usepackage{adjustbox}
\usepackage{multirow}
\usepackage{mathrsfs}
\usepackage{hyperref}
\usepackage{booktabs}
\usepackage[table]{xcolor}
\usepackage{threeparttable}
\usepackage{cleveref}

\begin{document}
\title{\textbf{DSCA}: A \textbf{D}igital Subtraction Angiography \textbf{S}equence Dataset and Spatio-Temporal Model for \textbf{C}erebral \textbf{A}rtery Segmentation}

\author{Jiong Zhang, Qihang Xie, Lei Mou, Dan Zhang, Da Chen, Caifeng Shan,\\ Yitian Zhao, Ruisheng Su, and Mengguo Guo
\thanks{This work was supported in part by the National Science Foundation Program of China (62371442), in part by the Zhejiang Provincial Natural Science Foundation (LR24F010002, LQ23F010002, LZ23F010002), in part by the Ningbo Natural Science Foundation (2022J143).}
\thanks{J. Zhang, Q. Xie, L. Mou, Y. Zhao are with the Laboratory of Advanced Theranostic Materials and Technology, Ningbo Institute of Materials Technology and Engineering, Chinese Academy of Sciences, Ningbo, China.}
\thanks{D. Zhang is with the School of Cyber Science and Engineering, Ningbo University of Technology, Ningbo, China.}
\thanks{D. Chen is with the Shandong Artificial Intelligence Institute, Qilu University of Technology, Jinan, China.} 
\thanks{C. Shan is with the School of Intelligence Science and Technology, Nanjing University, Nanjing, China.}
\thanks{R. Su is with the Erasmus MC, University Medical Center, Rotterdam, Netherlands. He is also with the Department of Biomedical Engineering, Eindhoven University of Technology, Eindhoven, Netherlands.}
\thanks{M. Guo is with the First Affiliated Hospital of Zhengzhou University, Zhengzhou, China.}
\thanks{Corresponding author: Mengguo Guo, Da Chen, and Qihang Xie (gmengguo86@gmail.com; dachen.cn@hotmail.com; xieqihang@nimte.ac.cn)}
}

\maketitle
\begin{abstract}
Cerebrovascular diseases (CVDs) remain a leading cause of global disability and mortality. Digital Subtraction Angiography (DSA) sequences, recognized as the gold standard for diagnosing CVDs, can clearly visualize the dynamic flow and reveal pathological conditions within the cerebrovasculature. Therefore, precise segmentation of cerebral arteries (CAs) and classification between their main trunks and branches are crucial for physicians to accurately quantify diseases. However, achieving accurate CA segmentation in DSA sequences remains a challenging task due to small vessels with low contrast, and ambiguity between vessels and residual skull structures. Moreover, the lack of publicly available datasets limits exploration in the field. In this paper, we introduce a DSA Sequence-based Cerebral Artery segmentation dataset (DSCA), the publicly accessible dataset designed specifically for pixel-level semantic segmentation of CAs. Additionally, we propose DSANet, a spatio-temporal network for CA segmentation in DSA sequences. Unlike existing DSA segmentation methods that focus only on a single frame, the proposed DSANet introduces a separate temporal encoding branch to capture dynamic vessel details across multiple frames. To enhance small vessel segmentation and improve vessel connectivity, we design a novel TemporalFormer module to capture global context and correlations among sequential frames. Furthermore, we develop a Spatio-Temporal Fusion (STF) module to effectively integrate spatial and temporal features from the encoder. Extensive experiments demonstrate that DSANet outperforms other state-of-the-art methods in CA segmentation, achieving a Dice of 0.9033.
\end{abstract}


\section{Introduction}
\label{sec:introduction}
Cerebrovascular diseases (CVDs) are significant causes of death, resulting in considerable suffering for patients and placing substantial financial pressure on families~\cite{roth2020global}. The majority of CVDs, such as stroke, moyamoya disease, and cerebral aneurysms, affect the geometrical and topological conditions of cerebral arteries (CAs). Therefore, segmenting CAs is essential for revealing vascular abnormalities associated with these diseases, leading to timely diagnosis and treatment planning~\cite{goni2022brain}. 


In clinical practice, several imaging techniques such as computed tomography angiography (CTA), magnetic resonance angiography (MRA), and digital subtraction angiography (DSA) are commonly used for the diagnosis and treatment of CVDs. Among them, DSA offers dynamic imaging of cerebral vessels with high spatial and temporal resolution, providing accurate details of lesions~\cite{hess2018imaging}. It is thus commonly used as the gold standard for diagnosing CVDs by clinicians, 
in cases where CTA and MRA fail to provide accurate diagnosis. 
However, the examination using DSA heavily relies on visual inspection by radiologists, which can be time-consuming and labor-intensive. Furthermore, radiologists with limited experience may be susceptible to misdiagnosis and underdiagnosis. Therefore, automated segmentation of CAs in DSA sequences, with the capability of capturing precise cerebrovascular details, offers valuable features to expedite the diagnosis and treatment.
\begin{figure*}[!t]
	\centering
        \includegraphics[width=\linewidth]{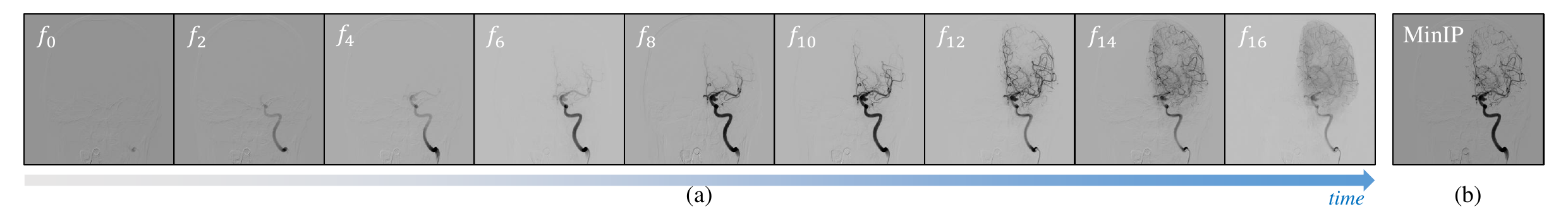}
	\caption{(a) is a DSA sequence, which includes 17 frames in the arterial phase. (b) is the MinIP image of the arterial phase of the DSA sequence.}
	\label{fig-1}
\end{figure*}

In terms of vascular segmentation, geometric-based and learning-based techniques have experienced rapid development and shown considerable promise in different modalities. Compared to traditional methods such as 
model-based~\cite{zou2008model} and tracking-based methods~\cite{sang2007knowledge}, deep learning-based techniques demonstrate superior robustness in cerebrovascular segmentation~\cite{chen2022generative,meng2020multiscale}. Notably, the UNet architecture~\cite{ronneberger2015u} emerges as an important milestone, with its diverse variants being widely adopted for both 2D and 3D segmentation tasks~\cite{fu2016deepvessel, liskowski2016segmenting, ma2020rose}. 
These models incorporate carefully designed components, such as improved skip connections~\cite{meng2020multiscale, liu2022full, wang2022uctransnet}, multi-scale module designs~\cite{meng2020multiscale, mou2019dense, ye2022mfi}, attention mechanisms~\cite{mou2021cs2, schlemper2019attention, zhang2019attention}, and transformer blocks~\cite{chen2021transunet, cao2022swin, hatamizadeh2022unetr}, aiming at augmenting segmentation efficacy. 
Besides that, certain studies explore aspects beyond network structure considerations. For instance, Isensee~\textit{et al.}~\cite{isensee2021nnu} introduced nnUNet, a UNet model incorporated with extensive data preprocessing and training strategies, which proved successful in numerous segmentation competitions. Zhang~\textit{et al.}~\cite{zhang2023centerline} improved coronary artery segmentation by utilizing a multi-task approach that prioritized centerline segmentation. In the field of cerebrovascular segmentation, some researchers explored the integration of Generative Adversarial Networks (GANs) to improve the segmentation performance~\cite{chen2022generative}, while others leverage semi-supervised and unsupervised methodologies to address challenges associated with data labeling~\cite{fan2020unsupervised,xie2022semi}.

Moccia~\textit{et al.}~\cite{moccia2018blood} and Mookiah~\textit{et al.}~\cite{mookiah2021review} summarize numerous vascular segmentation methods and datasets, which have significantly advanced the field of vessel segmentation. However, the majority of these methods are applied to retinal vessel segmentation. Compared to other imaging techniques, such as color fundus photography, acquiring DSA images is more challenging and involves higher risks. This has resulted in limited research on CA segmentation in DSA images. In DSA cerebrovascular segmentation, Zhang~\textit{et al.}~\cite{zhang2020neural} first proposed a straightforward U-shaped network for single-frame segmentation of DSA cerebral vessels. Patel~\textit{et al.}~\cite{patel2020multi} compared the performance of UNet~\cite{ronneberger2015u} and DeepMedic~\cite{kamnitsas2017efficient} in DSA cerebral vessel segmentation, demonstrating the effectiveness of DeepMedic's use of multi-resolution input. Meng~\textit{et al.}~\cite{meng2020multiscale} introduced the MDCNN framework, leveraging CNN with multi-scale dense connections in conjunction with an entropy-sampled patch method for single-frame DSA segmentation. Xu~\textit{et al.}~\cite{xu2023ernet} proposed an edge regularization network (ERNet) for segmenting vessels in DSA images. ERNet uses erosion and dilation on the initial binary annotations to create pseudo-ground-truths for false negative and false positive cases. Subsequently, Vepa~\textit{et al.}~\cite{vepa2022weakly} proposed a weakly supervised approach employing an active contour model to generate pseudo-labels for DSA vessel segmentation. However, despite these advancements, the segmentation performance of these methods still suffers from the low contrast of small vessels, and the interference from residual skull structures. Additionally, these methods focus only on pixel-by-pixel vessel extraction, while further classifying the vessel into bifurcated vessels (BV) and main artery trunk (MAT) is more clinically significant, e.g., for automatic detection and evaluation of stenosis, and automatic cerebrovasculature labeling \cite{chng2008territorial}. Moreover, regardless of whether fully supervised or weakly supervised methods are employed, existing approaches typically confine training on a single frame extracted from DSA sequences, leading to unstable microvascular segmentations.

In clinical practice, a DSA sequence displays dynamic flows of vessels, where each frame captures only a fraction of the contrast agent, as depicted in Fig. \ref{fig-1}-(a). Moreover, pathological changes such as arterial stenosis, collateralization, and aneurysms associated with various diseases exhibit significant variability across DSA sequences, appearing at uncertain frames within the sequence and manifesting diverse morphological characteristics. Consequently, relying solely on a single DSA frame for segmentation fails to cover the entire cerebrovasculature and these pathological features adequately.

As illustrated in Fig.~\ref{fig-1},  the minimum intensity projection (MinIP) image reveals the complete angiogram, incorporating information from all frames. These observations motivate us to consider training on entire DSA sequences and performing labeling on the MinIP image. This approach not only preserves the complete cerebrovascular features but also alleviates the annotation workload simultaneously. Moreover, our previous work~\cite{xie2024dsnet} has demonstrated that temporal information can enhance cerebrovascular segmentation performance. Recently, Liu~\textit{et al.}~\cite{liu2024dias} publicly released a DSA sequence dataset, DIAS. They established benchmarks for this dataset across three levels: fully supervised, weakly supervised, and semi-supervised, significantly advancing DSA research. The DIAS dataset contains 60 labeled samples and 60 unlabeled samples, and it lacks detailed descriptions of the pathological information.

Therefore, in this work, we establish a DSA Sequence dataset, comprising 224 DSA sequences of several common CVDs. To address the segmentation challenges, we propose a spatio-temporal network (DSANet) for CA segmentation. It takes a DSA sequence and its MinIP image as input to produce a complete 2D segmentation. Unlike 2D segmentation methods that only use a spatial encoding branch (SEB), our network includes an independent temporal encoding branch (TEB) to capture crucial temporal information. Following the TEB, we treat each frame in the sequence as a token, leveraging TemporalFormer blocks to capture global flow information from the DSA sequence. Subsequently, the spatial and temporal features from the two encoding branches are effectively fused through the spatio-temporal fusion (STF) module. 
In summary, our main contributions can be outlined as follows:

$\bullet$ We establish the new publicly accessible \textbf{D}SA-\textbf{S}equence based \textbf{C}erebral \textbf{A}rtery segmentation dataset (DSCA), consisting of pixel-wise cerebrovascular annotations across 224 DSA sequences from several common CVDs. To promote further developments in this field, the DSCA dataset, the code, and baseline models are made available at this website\footnote{\url{https://github.com/jiongzhang-john/DSCA}}.

$\bullet$ We propose a novel spatio-temporal strategy aimed at enhancing cerebral artery segmentation within DSA sequences. 
We design two encoding branches, i.e., the TEB and SEB, which respectively employ the sequence images and MinIP images as input, to leverage both spatial and temporal information for precise segmentation of the entire cerebrovasculature.

$\bullet$ To improve segmentation connectivity, we propose treating each frame as individual tokens following the TEB. These tokens are combined and fed into specially designed TemporalFormer blocks to effectively capture global context and dynamic correlations among frames across the sequence.

$\bullet$ We introduce a new STF module to integrate spatial and temporal features from the encoders. This module effectively addresses ambiguity among BV, MAT, and residual skulls by leveraging the dynamic changes of vessels and the structural invariance of the skull.



\begin{table} 
    \centering
    \caption{Dataset description. Internal Carotid Artery (ICA), External Carotid Artery (ECA), Vertebral Artery (VA), Antero-Posterior (AP), Lateral (LA). }
    \label{tab:dataset description}
    \begin{tabular}{c|c|c|c|c}
    \toprule
         \textbf{Artery} & \textbf{Total} & \textbf{View} &  \textbf{Devices} & \textbf{Diseases}\\ \midrule
         \multirow{4}{*}{ICA} & \multirow{4}{*}{126} & AP: 62 & Siemens:36 & Normal: 28\\
         & & & Philips: 84 & Stenosis: 23\\
         & & LA: 64 & Shimadzu: 6 & Moyamoya: 48\\ 
         &&&& Aneurysm: 27\\ \midrule
         \multirow{3}{*}{ECA} &  & AP: 26 & Siemens: 30 & \multirow{3}{*}{Normal: 55} \\
         & 55 &  & Philips: 22 & \\
         & & LA: 29 & Shimadzu: 3 & \\ \midrule
         \multirow{3}{*}{VA} &  & AP: 17 & Siemens: 17 & Normal: 37\\
         & 43 &  & Philips: 23 & Stenosis: 2\\
         & & LA: 26 & Shimadzu: 3 & Moyamoya: 4\\ \midrule
         \bottomrule
    \end{tabular} 
\end{table}

\section{Dataset}
\subsection{Dataset Description}
The DSCA dataset was collected from intraoperative and postoperative DSA imaging at the First Affiliated Hospital of Zhengzhou University, Zhengzhou, China. The acquisition of these subjects was retrospectively between January 2022 and July 2023, with patient identifiers removed to protect privacy. All the data used in this study were collected under the approval of institutional ethics committees and consented by the participants, following the Declaration of Helsinki.

The DSCA dataset consists of 58 patients, in total 224 DSA sequences including 1792 images from the left and right hemispheres. Among them, there are 28 male and 30 female patients, with ages ranging from 9 to 81 years and an average age of 49. Notably, the DSA sequences include three different arteries: the internal carotid artery (ICA), external carotid artery (ECA), and vertebral artery (VA). Specifically, there are 126 sequences for ICA, 55 for ECA, and 43 for VA. These sequences were captured by multiple imaging devices including AXION-Artis-160145 and AXION-Artis-160480 (Siemens, Germany), Azurion-559, Azurion-703844, AlluraXper-722012-2542 and AlluraXper-722038-129 (Philips, Netherlands), and Bransist Safire VC17 (Shimadzu, Japan), with sampling rates ranging from 4 to 7 frames per second,  which is slightly different from what is reported in~\cite{ahn2013basic}. The disease and manufacturer information of each sequence are shown in Table \ref{tab:dataset description}. Moreover, these sequences exhibit different resolutions (from $512\times472$ to $1432\times1432$). Each DSA sequence, captured in antero-posterior or lateral views, was stored in a DICOM file, with 108 DSA sequences in the antero-posterior view and 116 in the lateral view. All sequences retained arterial phase frames while discarding non-contrast, capillary phase, and venous phase frames~\cite{su2021autotici}, guided by the neurosurgeon's expertise. Therefore, DSA sequences in DSCA dataset consist of 5 to 22 frames. To align with network input, we resample all DSA sequences to 8 frames based on the distribution of extracted arterial phase frames. Detailed information about the dataset is provided in the GitHub\footnotemark[\value{footnote}].

\begin{table}
    \centering
    \caption{Dataset comparison}
    \label{tab:dataset}
    \begin{tabular}{p{2.3cm}<{\centering}p{0.7cm}<{\centering}ccp{0.7cm}<{\centering}}
    \toprule
         \textbf{Dataset} & \textbf{Sample} & \textbf{Resolution} & \textbf{Type} & \textbf{Public}\\ \midrule
         Meng \textit{et al.}~\cite{meng2020multiscale} & 30 & 700 $\times$ 700 & Single Frame & No\\
         Zhang \textit{et al.}~\cite{zhang2020neural} & 20  & 512 $\times$ 512 & Single Frame & No\\
         Vepa \textit{et al.}~\cite{vepa2022weakly} & 128 & 1024 $\times 1024$ & Single Frame & No\\
         Fu \textit{et al.}~\cite{fu2016vessel} & 88 & 1024 $\times$ 1024 & Single Frame & No\\
         Xu \textit{et al.}~\cite{xu2023ernet} & 138 & 800 $\times$ 800 & Single Frame & No\\
         Patel \textit{et al.}~\cite{patel2020multi} & 100 & - & Single Frame & No\\
         DIAS~\cite{liu2024dias} & 120 & 800 $\times$ 800 & Sequence & Yes \\ 
         \midrule
         \textbf{DSCA} & \textbf{224} & \textbf{Multiple} & \textbf{Sequence} & \textbf{Yes}\\ 
         \bottomrule
    \end{tabular}
\end{table}

\begin{figure}[!t]
	\centering
	\includegraphics[width=\linewidth]{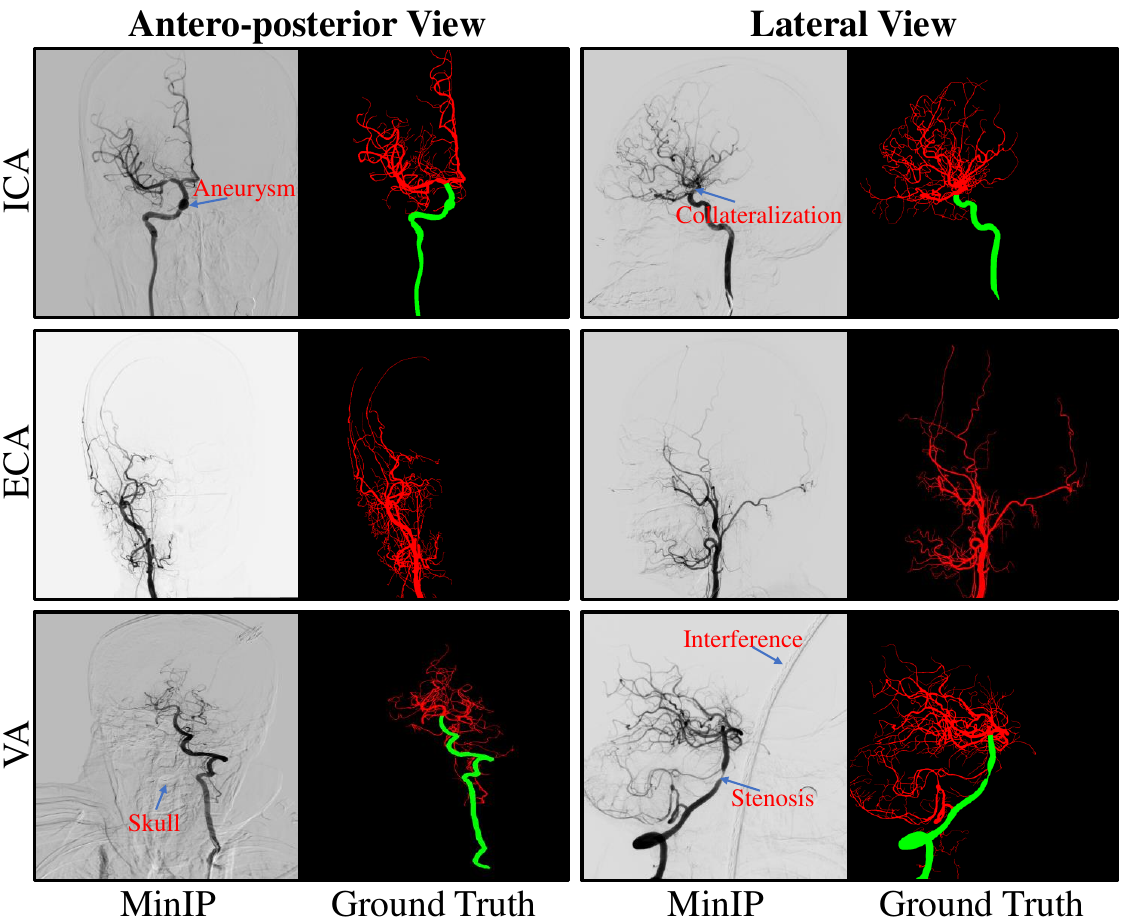}
	\caption{The annotating examples of the DSCA dataset. The first to third rows are ICA, ECA, and VA. The 1st and 2nd columns are antero-posterior view, and the 3rd and 4th columns are lateral view.}
	\label{fig-label}
\end{figure}

\subsection{Manual Annotation Protocol}
As mentioned above, manually annotating the complete DSA sequence is a time-consuming and labor-intensive task. In contrast, annotating the corresponding MinIP image can save considerable time while retaining the entire cerebrovascular information. Furthermore, distinguishing between BV and MAT holds greater clinical significance. Thus, we asked skilled clinicians to manually annotate the MinIP images, including specific labeling for both BV and MAT. These annotations served as the gold standard for the DSCA dataset. The data preprocessing and annotation procedures are as follows: (1) Selection of arterial phase frames from the DSA sequences. (2) Registration of the extracted arterial phase sequence frames using Elastix~\cite{klein2009elastix}, followed by resampling to 8 frames. (3) Projection of the resampled frames to generate MinIP images. (4) Annotation of the MinIP images utilizing ITK-SNAP~\cite{yushkevich2016itk}.

The annotation process of the DSCA dataset was performed by five skilled clinicians with over 5 years of experience who received complete training. They were evenly assigned to annotate the 224 MinIP images, each being annotated only once. Disputed annotations were referred to a senior neurosurgeon for confirmation. Subsequently, a second senior neurosurgeon conducted a final review. If there were different opinions, the two neurosurgeons worked together to reach a consensus. The annotated results of the DSCA dataset are presented in Fig. \ref{fig-label}, wherein the ICA and VA sequences were categorized into background, BV, and MAT. The ECA sequences, lacking a distinct main trunk, were classified into background and BV. The criteria for dividing MAT and BV are based on the last segment of the ICA and VA. The ICA is labeled up to the C7 segment, and the VA up to the V4 segment, merging with the basilar artery. To mitigate potential risks of data leakage, the DSA sequences were randomly divided based on patient IDs into a training set comprising $80\%$ of the data and a test set comprising the remaining $20\%$. This ensures that the same patient cannot appear in both the training and testing sets simultaneously. Consequently, the training set comprises 180 sequences, while the test set contains 44 sequences. 

Additionally, Table~\ref{tab:dataset} provides a detailed comparison of the DSCA dataset and other DSA datasets. Notably, the DSCA dataset offers several distinct advantages, including a large number of samples, diverse image resolutions, and most importantly, public accessibility. Even compared to the recently released DIAS dataset, which is also a DSA sequence dataset, DSCA shows clear advantages over DIAS in terms of sample size and diversity in image resolution.

\begin{figure*}[t]
	\centering	\includegraphics[width=\textwidth]{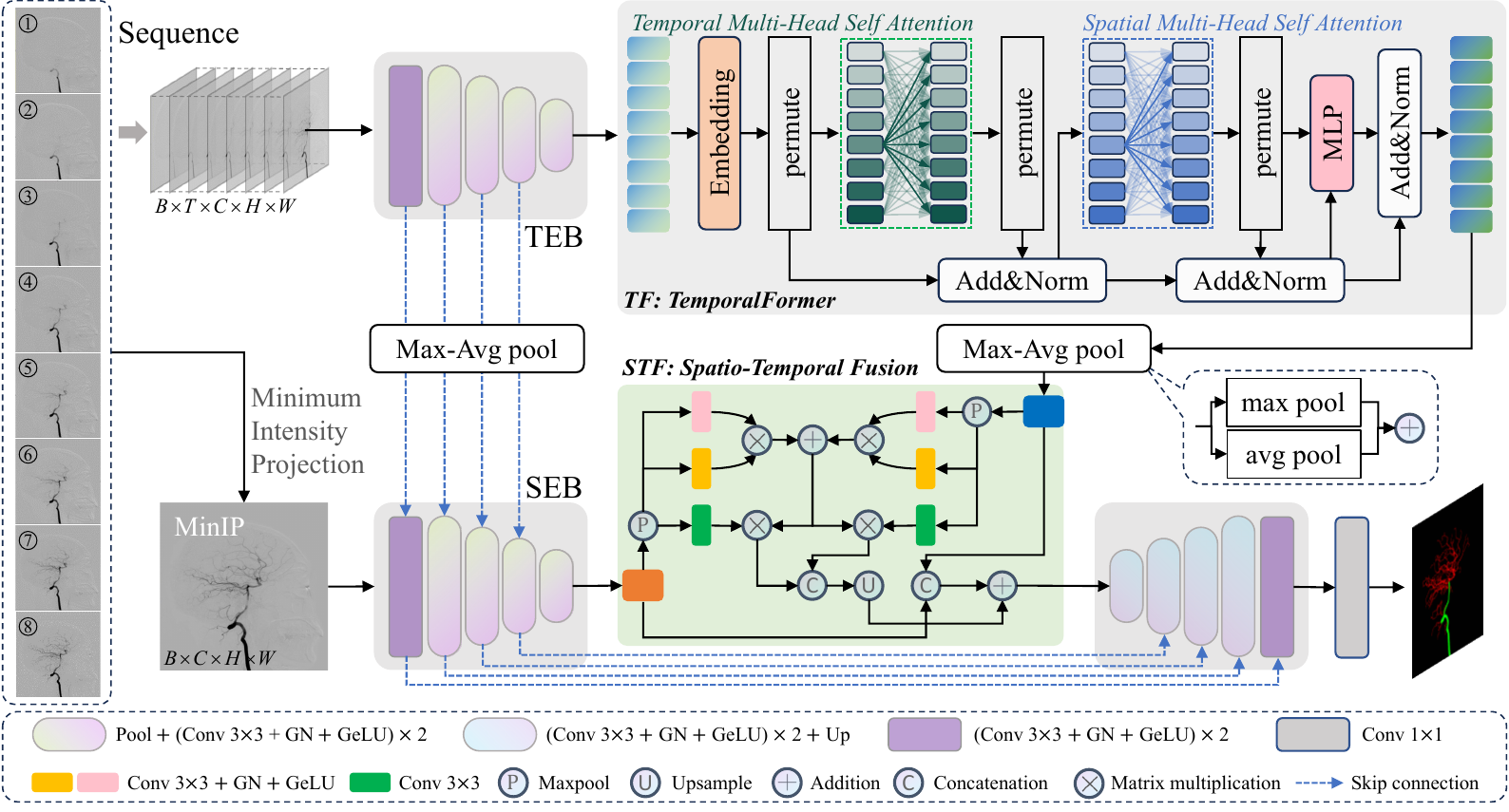}
	\caption{Architecture of our proposed DSANet. It is primarily composed of two encoding branches, one decoding branch, and two modules: \textbf{T}emporal\textbf{F}ormer (TF), \textbf{S}patio-\textbf{T}emporal \textbf{F}usion (STF). The details of the dimension transformation process of TF are shown in Fig. 4}
	\label{fig-framework}
\end{figure*}

\section{Proposed method}
In this section, we introduce DSANet, a spatio-temporal network designed to segment CA in DSA sequences. The DSANet architecture comprises two encoding branches: the SEB and the TEB, along with a decoding branch, as depicted in Fig.~\ref{fig-framework}. This network incorporates an additional TEB to extract temporal information from DSA sequences. After passing through the TEB, each sequential frame is treated as a token, and the TemporalFormer is developed to capture the global context and significant correlations among frames. In the bottleneck, an STF module is designed to fuse spatial and temporal features derived from different encoding branches.

\subsection{Encoding Branches}
As shown in Fig. \ref{fig-framework}, there are two encoding branches: TEB and SEB. Although structurally similar, these branches function independently without weight sharing. Each branch adopts a five-layer structure, comprising a convolutional layer with double $3\times3$ convolutional blocks, double GroupNorm blocks, and double Gelu blocks, followed by four downsampling layers. The downsampling layer comprises a Maxpool block and a convolutional layer. The primary difference between the two encoding branches lies in their input.
\subsubsection{Spatial encoding branch}
In the SEB, the input is a 2D MinIP image $M \in \mathbb{R}^{B \times C \times H \times W}$, where $B$, $C$, $H$, and $W$ represent the batch size, number of channels, height, and width of $M$. And $C$ is equal to 1. This input setting is consistent with other 2D segmentation tasks in our subsequent comparisons.
\subsubsection{Temporal encoding branch}
In the TEB, 
the input is organized in 2D sequential frames denoted as $S \in \mathbb{R}^{B \times T \times C \times H \times W}$, where $T$ represents the number of frames. The sequence encoding process is optimized for training convenience by merging $B$ and $T$. 
Following the first convolutional layer, the value of $C$ is doubled. Afterward, as the input sequence passes through the downsampling layer, both $H$ and $W$ are halved, and $C$ is doubled. Each of the first four layers of the encoding branches includes an additional output serving as a skip connection. In the TEB, the skip connection needs to be downsampled in the $T$ dimension and then concatenated with the skip connection in the SEB to form the final skip connection, and is subsequently fed into the corresponding layer in the decoder.

\subsection{TemporalFormer}
The dynamic flows within DSA sequences possess rich temporal information, characterized by meaningful contextual correlations among frames. Consequently, this temporal information offers valuable assistance in segmenting cerebral arteries and categorizing them into BV and MAT.
To effectively capture the global context and inter-frame correlations, we feed the features generated by each frame in the TEB into the TemporalFormer (TF), as depicted in Fig.~\ref{fig-framework}. 

\begin{figure*}[!t]
	\centering
	\includegraphics[width=\linewidth]{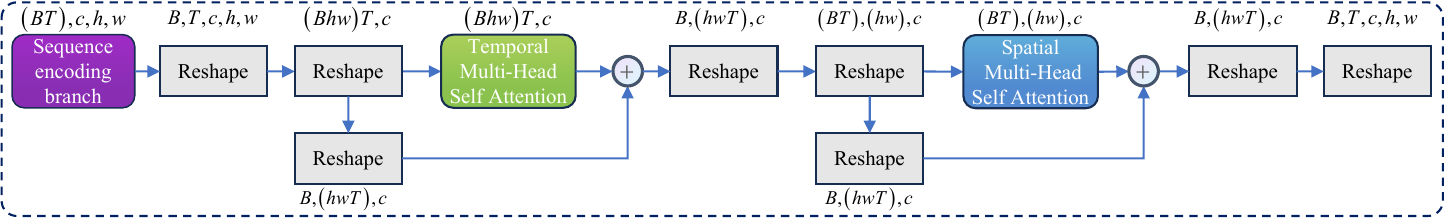}
	\caption{The details of the dimension transformation process in TemporalFormer, where c is 512.}
	\label{fig-dimension}
\end{figure*}

We consider the features $F_s \in \mathbb{R}^{(BT) \times c \times h \times w}$ extracted from the TEB, where each frame feature is defined as $f^t_s \in \mathbb{R}^{B \times c \times h \times w}$. Here, $c$, $h$, and $w$ respectively represent the number of channels, the height, and the width of $F_s$. The index $t \in \{0, T\}$ is treated as a token and inputted into the TF. Specifically, $F_s$ is initially reshaped into $\in \mathbb{R}^{B \times T \times c \times h \times w}$ to restore its original dimensions. Then, it is converted to a feature vector $\mathcal{F}_s$ $\in \mathbb{R}^{(Bhw) \times T \times c}$ by flattening $B$, $h$, and $w$. Subsequently, we integrate learnable position encoding $f_p \in \mathbb{R}^{T \times c }$ with $\mathcal{F}_s$ by adding them together. Following this, $\mathcal{F}_s$ is transformed into $\in \mathbb{R}^{B \times (hwT) \times c}$ to generate a residual feature denoted as $\mathcal{F}^{\prime}_s$, which is then restored. Afterwards, $\mathcal{F}_s$ undergoes a temporal multi-head self-attention operation, which can be formulated as:
\begin{equation}\label{1}
    \begin{aligned}
    q_{s}, k_{s}, v_{s} = \mathbb{W}(\mathcal{F}_s),
    \end{aligned}
\end{equation}
\begin{equation}\label{2}
    \begin{aligned}
    \tilde{\mathcal{F}}_{s} = Softmax(\frac{q_{s} \times k_{s}^{\mathtt{T}}}{\sqrt{d_k}})v_{s},
    \end{aligned}
\end{equation}
where $q_{s}, k_{s}, v_{s}, \tilde{\mathcal{F}}_{s} \in \mathbb{R}^{(Bhw) \times T \times c }$ and $d_k$ is the dimension of $k_{s}$, and $\mathbb{W}$ represents a linear function. The output $\tilde{\mathcal{F}}_{s}$ from the temporal self-attention block is reshaped to $\in \mathbb{R}^{B \times (hwT) \times c}$. This reshaped output is then added to $\mathcal{F}^{\prime}_s$ to obtain a new $\tilde{\mathcal{F}}_{s}$ and $\tilde{\mathcal{F}}^{\prime}_{s}$, where $\tilde{\mathcal{F}}^{\prime}_{s}$ is equivalent to $\tilde{\mathcal{F}}_{s}$. Next, we use a rearrange operation to map the new feature vector $\tilde{\mathcal{F}}_{s}$ into $\in \mathbb{R}^{(BT) \times (hw) \times c}$. Similarly, a self-attention mechanism is applied in the spatial domain, as depicted in Equation \eqref{1} and \eqref{2}. Given $\hat{\mathcal{F}}_{s}$ as the output of the spatial self-attention block, we reshape it into $\mathbb{R}^{B \times (hwT) \times c}$ and add it to $\tilde{\mathcal{F}}^{\prime}_{s}$ to get a new enhanced $\hat{\mathcal{F}}_{s}$. Following these self-attention blocks, an MLP block with two linear layers is applied to $\hat{\mathcal{F}}_{s}$. The details of the dimension transformation process are shown in Fig. \ref{fig-dimension}.
After passing through four TF layers, we finally obtain the enhanced feature $\hat{\mathcal{F}}_{s}$ $\in \mathbb{R}^{B \times (hwT) \times c}$, and we restore it to $\hat{F}_s$ $\in \mathbb{R}^{B \times T \times c \times h \times w}$. By downsampling $\hat{F}_s$ in the $T$ dimension, we get the new feature $F_s \in \mathbb{R}^{B \times c \times h \times w}$.


\subsection{Spatio-Temporal Fusion Module}
To effectively fuse the spatial and temporal features from the encoder, we introduce an STF module. Given the feature $F_m \in \mathbb{R}^{B \times c \times h \times w}$ from the SEB and the feature $F_s \in \mathbb{R}^{B \times c \times h \times w}$ from the TF module, we initially apply Maxpool to both $F_m$ and $F_s$ to reduce the size of feature maps and reduce noise interference to retain critical features, yielding $F^{\prime}_m$, and $F^{\prime}_s$ $\in \mathbb{R}^{B \times c \times \frac{h}{2} \times \frac{w}{2}}$. Subsequently, the two features are passed through a shared convolutional block to generate feature vector $Q_{m}, K_{m}, V_{m}, Q_{s}, K_{s}, V_{s}$ $\in \mathbb{R}^{B \times c \times \frac{hw}{4}}$. Afterward, $Q_{m}, Q_{s}$ are multiplied by $K_{m}^{\mathtt{T}}, K_{s}^{\mathtt{T}}$, respectively, and then dividing by $\sqrt{c}$ to prevent the dot product from becoming too large and causing the vanishing gradient problem. The weight matrices $\alpha_i$, $\alpha_s$ $\in \mathbb{R}^{B \times \frac{hw}{4}}$ are obtained through the softmax function, similar to Equation \eqref{2}. Additionally, we believe that spatio-temporal features at the same position can enhance the representation of the network. Therefore, the two weight matrices are then summed, and the result is respectively multiplied by $V_{m}$ and $V_{s}$ to get the enhanced features $\tilde{F}^{\prime}_m$, $\tilde{F}^{\prime}_s$ $\in \mathbb{R}^{B \times \frac{hw}{4}}$, as depicted below: 
\begin{equation}\label{5}
\begin{aligned}
    \tilde{F}^{\prime}_m = (\alpha_i + \alpha_s) \times V_{m},\;\;
    \tilde{F}^{\prime}_s = (\alpha_i + \alpha_s) \times V_{s}.
\end{aligned}
\end{equation}
We concatenate $\tilde{F}^{\prime}_m$, $\tilde{F}^{\prime}_s$, and use an upsample and reshape operation to map the result back to the standard feature map $\tilde{F}^{\prime}_{sm}$ $\in \mathbb{R}^{B \times 2c \times h \times w}$. Finally, $F_m$ and $F_s$ are concatenated and added with $\tilde{F}^{\prime}_{sm}$ in an element-wise manner to obtain the refined feature $\hat{F}_{sm}$, as defined by:
\begin{equation}\label{6}
\begin{aligned}
    \hat{F}_{sm} = \tilde{F}^{\prime}_{sm} + Concat(F_s, F_m),
\end{aligned}
\end{equation}
where $Concat$ indicates the concatenation operation. The feature $\hat{F}_{sm}$ is then fed into the decoder for feature restoration.

\subsection{Loss Function}
We utilize the cross-entropy loss ($\mathcal{L}_{ce}$) and the dice loss ($\mathcal{L}_{dice}$) to optimize the DSANet. The deep supervision mechanism \cite{isensee2021nnu} is also deployed to regularize the training process further. 
The $\mathcal{L}_{ce}$ is defined by:
\begin{equation}\label{7}
\begin{aligned}
    \mathcal{L}_{ce} = \frac{1}{N} \sum_j \mathcal{L}_j = -\frac{1}{N}\sum_j\sum^{M}_{k=1}y_{jk}log(p_{jk}),
\end{aligned}
\end{equation}
where $N$ represents the total number of pixels, $M$ gives the number of classes, $y_{jk} \in \{0, 1\}$ defines whether pixel $j$ belongs class $k$ in the GT, and $p_{jk} \in [0, 1]$ indicates the probability that pixel $j$ belongs to class $k$ in the predicted map. 
Similarly, $\mathcal{L}_{dice}$ can be defined as:
\begin{equation}\label{8}
\begin{aligned}
    \mathcal{L}_{dice} = 1 - \frac{2 \sum^{N}_{j=1} p_j \times y_j}{\sum^{N}_{j=1}p_j + \sum^{N}_{j=1}y_j}.
\end{aligned}
\end{equation}
Therefore, the overall loss function is given by:
\begin{equation}\label{9}
\begin{aligned}
    \mathcal{L}_{total} = \sum^2_{a=0} \frac{1}{2^{a}}(\mathcal{L}^{a}_{ce} + \mathcal{L}^{a}_{dice}),
\end{aligned}
\end{equation}
where $a$ is the downsampling scale of the full-size output.

\section{Experiments}
\subsection{Implementation Details}
The proposed method was implemented with Python3.7, PyTorch 1.11.0, and Ubuntu 18.04. All training procedures were performed on RTX3090 (24G) GPUs. We used nnUNet as the backbone network. 
Stochastic Gradient Descent (SGD) was employed for model optimization during training, with a momentum of 0.99 and a weight decay of 3e-5. The initial learning rate was set to 0.01 with a polynomial decay strategy.  The maximum epoch number was set to 500 and one epoch contains 100 iterations for fair comparisons. 

The preprocessing pipelines vary between training and testing. For training, we utilized a five-fold cross-validation method by dividing the 180 sequences in the training set, with four folds for training and one for validation. 
Each batch comprised 2 samples at 512$\times$512 pixels which were randomly cropped from images of varying dimensions. Both our method and the comparison methods used data pre-processed by nnUNet, and no post-processing was applied as it may result in the removal of small vessels. This post-processing method, which discards fractured small vessels, is unreasonable and may adversely affect the performance of comparative experiments, especially the clDice~\cite{shit2021cldice} metric. Additionally, common data augmentation methods were applied to optimize the training process, including random rotation, elastic deformation, random scaling, random cropping, gamma augmentation, and mirroring. For testing, we only perform mirroring operations for data augmentation. The testing procedure involved cropped patches of 512$\times$512 pixels from the test image via sliding windows. We applied padding with mirroring to ensure consistency in patch sizes. Subsequently, these patches were inputted into the network for prediction and then sequentially reconstructed to match the original image size.
Finally, we compared the reconstructed predictions with the GT across the entire image. The performance metrics on the test set were obtained by averaging the quantitative segmentation results obtained from the five models based on the five-fold cross-validation.

\subsection{Evaluation Metrics}
To comprehensively evaluate the segmentation performance, the following metrics are calculated and compared: Jaccard (JAC)=TP/(TP+FP+FN); Dice=2$\times$TP/(2$\times$TP+FP+FN); Sensitivity (SEN)=TP/(TP+FN); Precision (PRE)=TP/(TP+FP); Area Under the ROC Curve (AUC), where TP, FP, TN, and FN are true positive, false positive, true negative, and false negative, respectively. The clDice~\cite{shit2021cldice} is also used to evaluate vessel connectivity when measuring the entire cerebral artery. Moreover, a paired t-test is conducted to calculate the p-values.

\begin{table*}[!t]
    \centering
    \caption{Quantitative comparisons with DSA segmentation and general segmentation methods on DSCA for BV and MAT. The results are average$\pm$standard deviation obtained from the five-fold cross-validation. All metrics are expressed as percentages (\%).}
    \label{tab:comparative-experiment}
    \resizebox{\textwidth}{!}{
    \renewcommand{\arraystretch}{1.25}
    \begin{tabular}{cl|cccccccccc}
    \toprule
         \multicolumn{2}{c}{\multirow{2}{*}{Methods}} \vline & \multicolumn{2}{c}{JAC} & \multicolumn{2}{c}{Dice} & \multicolumn{2}{c}{SEN} & \multicolumn{2}{c}{PRE} & \multicolumn{2}{c}{AUC} \\ \cmidrule{3-12}
         ~ & ~ & BV & MAT & BV & MAT & BV & MAT & BV & MAT & BV & MAT\\ \midrule
         \multicolumn{1}{c|}{\multirow{3}{*}{\rotatebox{90}{DSA}}}  & Zhang \textit{et al.}~\cite{zhang2020neural} & 67.56 $\pm$ 0.82 & 67.45 $\pm$ 1.36 & 80.11 $\pm$ 0.61 & 73.00 $\pm$ 1.26 & 76.70 $\pm$ 0.67 & 71.74 $\pm$ 1.46 & 84.61 $\pm$ 0.77 & 75.9 $\pm$ 1.09 & 88.08 $\pm$ 0.34 & 77.20 $\pm$ 1.31  \\
         \multicolumn{1}{c|}{} & MDCNN~\cite{meng2020multiscale} & 66.16 $\pm$ 2.95 & 69.57 $\pm$ 0.64 & 78.95 $\pm$ 2.23 & 74.61 $\pm$ 0.43 & 74.05 $\pm$ 4.27 & 74.12 $\pm$ 1.56 & 85.69 $\pm$ 1.15 & 77.09 $\pm$ 1.38 & 86.79 $\pm$ 2.10 & 79.06 $\pm$ 0.77   \\
         \multicolumn{1}{c|}{} & ERNet~\cite{xu2023ernet} & 73.39 $\pm$ 0.67 & 78.01 $\pm$ 1.07 & 84.19 $\pm$ 0.46 & 82.61 $\pm$ 1.18 & 81.67 $\pm$ 1.56 & 80.75 $\pm$ 1.50 & 87.49 $\pm$ 1.01 & 87.80 $\pm$ 2.02 & 90.60 $\pm$ 0.76 & 86.03 $\pm$ 1.54  \\ \midrule
         \multicolumn{1}{c|}{\multirow{9}{*}{\rotatebox{90}{General}}} & UNet~\cite{ronneberger2015u} & 71.30 $\pm$ 0.04 & 73.24 $\pm$ 1.76 & 82.74 $\pm$ 0.33 & 77.68 $\pm$ 1.67 & 80.10 $\pm$ 0.39 & 77.24 $\pm$ 1.69 & 86.24 $\pm$ 0.30 & 80.18 $\pm$ 2.59 & 89.80 $\pm$ 0.19 & 81.77 $\pm$ 1.62 \\ 
         \multicolumn{1}{c|}{} & CENet~\cite{gu2019net} & 68.34 $\pm$ 0.65 & 79.97 $\pm$ 2.59 & 80.69 $\pm$ 0.51 & 84.32 $\pm$ 2.78 & 76.86 $\pm$ 2.04 & 82.85 $\pm$ 2.73 & 85.78 $\pm$ 2.16 & 87.20 $\pm$ 2.67 & 88.18 $\pm$ 0.97 & 87.09 $\pm$ 2.52 \\ 
         \multicolumn{1}{c|}{} & CSNet~\cite{mou2021cs2} & 73.02 $\pm$ 0.93 & 81.28 $\pm$ 1.87 & 83.95 $\pm$ 0.65 & 85.37 $\pm$ 1.79 & 80.77 $\pm$ 2.05 & 84.11 $\pm$ 2.07 & 87.98 $\pm$ 1.23 & 89.13 $\pm$ 1.63 & 90.17 $\pm$ 0.99 & 88.85 $\pm$ 2.24 \\ 
         \multicolumn{1}{c|}{} & AttUNet~\cite{zhang2019attention} & 69.88 $\pm$ 0.78 & 74.58 $\pm$ 1.00 & 81.73 $\pm$ 0.57 & 79.10 $\pm$ 0.92 & 76.51 $\pm$ 1.59 & 77.85 $\pm$ 1.26 & 88.47 $\pm$ 1.12 & 82.68 $\pm$ 1.01 & 88.07 $\pm$ 0.77 & 82.31 $\pm$ 1.13 \\ 
         \multicolumn{1}{c|}{} & SwinUNet~\cite{cao2022swin} & 64.22 $\pm$ 0.44 & 67.41 $\pm$ 1.45 & 77.61 $\pm$ 0.35 & 73.45 $\pm$ 1.30 & 75.33 $\pm$ 0.50 & 71.92 $\pm$ 1.28 & 79.92 $\pm$ 1.69 & 75.79 $\pm$ 1.49 & 87.33 $\pm$ 0.25 & 77.28 $\pm$ 1.22 \\ 
         \multicolumn{1}{c|}{} & MISSFormer~\cite{huang2022missformer} & 70.09 $\pm$ 0.37 & 78.97 $\pm$ 3.67 & 81.89 $\pm$ 0.21 & 84.06 $\pm$ 3.37 & 79.10 $\pm$ 0.62 & 82.41 $\pm$ 3.39 & 85.58 $\pm$ 0.91 & 86.87 $\pm$ 3.54 & 89.30 $\pm$ 0.30 & 87.08 $\pm$ 3.19 \\ 
         \multicolumn{1}{c|}{} & H2Former~\cite{he2023h2former} & 69.47 $\pm$ 0.62 & 83.90 $\pm$ 3.79 & 81.47 $\pm$ 0.43 & 88.34 $\pm$ 3.84 & 77.85 $\pm$ 1.54 & 86.87 $\pm$ 3.62 & 86.21 $\pm$ 1.17 & 92.51 $\pm$ 4.42 & 88.67 $\pm$ 0.74 & 91.59 $\pm$ 3.82 \\ 
         \multicolumn{1}{c|}{} & TransUNet~\cite{chen2021transunet} & 72.43 $\pm$ 0.46 & 81.38 $\pm$ 1.24 & 83.63 $\pm$ 0.34 & 85.09 $\pm$ 1.19 & 80.99 $\pm$ 0.92 & 83.89 $\pm$ 1.12 & 87.20 $\pm$ 0.45 & 88.33 $\pm$ 2.08 & 90.24 $\pm$ 0.45 & 88.29 $\pm$ 1.48 \\ 
         \multicolumn{1}{c|}{} & nnUNet~\cite{isensee2021nnu} & 76.76 $\pm$ 0.45 & 82.54 $\pm$ 2.42 & 86.45 $\pm$ 0.31 & 86.22 $\pm$ 2.50 & 85.29 $\pm$ 0.45 & 85.70 $\pm$ 2.50 & 88.06 $\pm$ 0.59 & 88.08 $\pm$ 2.20 & 92.41 $\pm$ 0.22 & 89.87 $\pm$ 2.51 \\ \midrule
         \multicolumn{1}{c|}{\multirow{9}{*}{\rotatebox{90}{Sequence + MinIP}}} & nnUNet\textsuperscript{*}~\cite{isensee2021nnu} & 77.52 $\pm$ 0.22 & 85.90 $\pm$ 2.10 & 86.94 $\pm$ 0.17 & 89.88 $\pm$ 2.29 & 85.73 $\pm$ 0.32 & 88.76 $\pm$ 2.09 & 88.56 $\pm$ 0.31 & 91.93 $\pm$ 2.57 & 92.64 $\pm$ 0.15 & 93.00 $\pm$ 1.88\\
         \multicolumn{1}{c|}{} & UNet\textsuperscript{*}~\cite{ronneberger2015u} & 62.12 $\pm$ 1.17 & 56.67 $\pm$ 5.80 & 75.97 $\pm$ 0.86 & 64.74 $\pm$ 4.48 & 84.39 $\pm$ 2.53 & 89.24 $\pm$ 6.32 & 70.78 $\pm$ 2.87 & 75.26 $\pm$ 2.21 & 91.48 $\pm$ 1.16 & 70.50 $\pm$ 3.16\\
         \multicolumn{1}{c|}{} & MISSFormer\textsuperscript{*}~\cite{huang2022missformer} & 65.55 $\pm$ 1.98 & 70.69 $\pm$ 3.53 & 78.69 $\pm$ 1.49 & 77.09 $\pm$ 3.18 & \textbf{87.95} $\pm$ \textbf{0.69} & 73.89 $\pm$ 3.99 & 72.28 $\pm$ 2.68 & 83.34 $\pm$ 4.49 & \textbf{93.29} $\pm$ \textbf{0.25} & 82.82 $\pm$ 3.82\\
         \multicolumn{1}{c|}{} & H2Former\textsuperscript{*}~\cite{he2023h2former} & 63.55 $\pm$ 1.13 & 60.23 $\pm$ 5.53 & 77.14 $\pm$ 0.87 & 67.91 $\pm$ 5.28 & 84.68 $\pm$ 0.39 & 61.60 $\pm$ 5.77 & 72.05 $\pm$ 1.38 & 87.98 $\pm$ 1.34 & 91.66 $\pm$ 0.18 & 78.51 $\pm$ 2.98\\
         \multicolumn{1}{c|}{} & DeepMedic~\cite{kamnitsas2017efficient} & 60.37 $\pm$ 1.42 & 58.76 $\pm$ 1.68 & 74.46 $\pm$ 1.01 & 67.76 $\pm$ 1.02 & 72.08 $\pm$ 3.35 & 74.48 $\pm$ 1.93 & 78.63 $\pm$ 3.60 & 63.49 $\pm$ 3.26 & 85.66 $\pm$ 1.57 & 77.99 $\pm$ 0.93\\
         \multicolumn{1}{c|}{} &
         DSNet~\cite{xie2024dsnet} & 76.62 $\pm$ 0.32 & 78.82 $\pm$ 2.59 & 86.32 $\pm$ 0.23 & 82.76 $\pm$ 2.68 & 84.85 $\pm$ 0.53 & 81.82 $\pm$ 2.68 & 88.23 $\pm$ 3.25 & 85.25 $\pm$ 3.30 & 92.20 $\pm$ 2.59 & 85.88 $\pm$ 2.83 \\ 
         \multicolumn{1}{c|}{} &
         CAVE~\cite{su2024cave} & 69.11 $\pm$ 2.80 & 79.09 $\pm$ 2.16 & 81.12 $\pm$ 1.88 & 83.70 $\pm$ 2.15 & 83.26 $\pm$ 4.51 & 81.90 $\pm$ 2.28 & 80.48 $\pm$ 1.60 & 88.07 $\pm$ 2.43 & 91.22 $\pm$ 2.21 & 86.84 $\pm$ 1.91 \\ 
         \multicolumn{1}{c|}{} &
         VSSNet~\cite{liu2024dias} & 66.79 $\pm$ 0.62 & 60.87 $\pm$ 3.14 & 79.35 $\pm$ 0.48 & 67.91 $\pm$ 2.29 & 87.39 $\pm$ 1.22 & 65.90 $\pm$ 3.81 & 73.84 $\pm$ 1.17 & 73.14 $\pm$ 0.77 & 93.07 $\pm$ 0.58 & 74.95 $\pm$ 1.90 \\ 
         \multicolumn{1}{c|}{} & DSANet & \textbf{78.09} $\pm$ \textbf{0.20} & \textbf{88.22} $\pm$ \textbf{1.52} & \textbf{87.32} $\pm$ \textbf{0.13} & \textbf{92.26} $\pm$ \textbf{1.56} & 86.40 $\pm$ 0.39 & \textbf{91.37} $\pm$ \textbf{1.57} & \textbf{88.59} $\pm$ \textbf{0.53} & \textbf{94.01} $\pm$ \textbf{1.40} & 92.98 $\pm$ 0.18 & \textbf{94.98} $\pm$ \textbf{1.39} \\
         \bottomrule
    \end{tabular} }
    \begin{tablenotes}
    \item '*' represents that the model input is a sequence and its MinIP.
    \end{tablenotes}
\end{table*}

\begin{table*}
    \centering
    \caption{Quantitative comparisons with DSA segmentation and general segmentation methods on DSCA for the entire vessel. The results are average$\pm$standard deviation obtained from five-fold cross-validation. All metrics are based on percentages (\%).}
    \label{tab:comparative_all}
    \begin{tabular}{cl|cccccc}
    \toprule
        \multicolumn{2}{c}{Methods} \vline & clDice & JAC & Dice & SEN & PRE & AUC \\ \midrule
        \multicolumn{1}{c|}{\multirow{3}{*}{\rotatebox{90}{DSA}}} & Zhang \textit{et al.}~\cite{zhang2020neural} & 81.87 $\pm$ 0.85 & 76.63 $\pm$ 0.66 & 86.49 $\pm$ 0.44 & 83.71 $\pm$ 0.54 & 89.92 $\pm$ 0.61 & 91.62 $\pm$ 0.27 \\
        \multicolumn{1}{c|}{} & MDCNN~\cite{meng2020multiscale} & 80.00 $\pm$ 2.85 & 75.19 $\pm$ 2.17 & 85.50 $\pm$ 1.44 & 81.58 $\pm$ 3.12 & 90.61 $\pm$ 0.99 & 90.58 $\pm$ 1.53 \\ 
        \multicolumn{1}{c|}{} & ERNet~\cite{xu2023ernet} & 85.36 $\pm$ 0.39 & 79.65 $\pm$ 0.36 & 88.40 $\pm$ 0.23 & 85.75 $\pm$ 1.15 & 91.66 $\pm$ 0.90 & 92.68 $\pm$ 0.55 \\ \midrule
        \multicolumn{1}{c|}{\multirow{9}{*}{\rotatebox{90}{General}}} & UNet~\cite{ronneberger2015u} & 84.34 $\pm$ 0.29 & 78.41 $\pm$ 0.35 & 87.61 $\pm$ 0.23 & 85.47 $\pm$ 0.36 & 90.27 $\pm$ 0.22 & 92.51 $\pm$ 0.18 \\
        \multicolumn{1}{c|}{} & CENet~\cite{gu2019net} & 81.01 $\pm$ 0.40 & 75.06 $\pm$ 0.44 & 85.45 $\pm$ 0.31 & 81.79 $\pm$ 1.70 & 90.04 $\pm$ 1.68 & 90.66 $\pm$ 0.80 \\
        \multicolumn{1}{c|}{} & CSNet~\cite{mou2021cs2} & 84.84 $\pm$ 0.82 & 79.06 $\pm$ 0.71 & 88.07 $\pm$ 0.45 & 84.97 $\pm$ 1.63 & 91.71 $\pm$ 0.11 & 92.29 $\pm$ 0.79\\
        \multicolumn{1}{c|}{} & AttUNet~\cite{zhang2019attention} & 82.63 $\pm$ 0.52 & 76.77 $\pm$ 0.46 & 86.56 $\pm$ 0.29 & 82.10 $\pm$ 1.12 & \textbf{92.09} $\pm$ \textbf{0.88} & 90.86 $\pm$ 0.54 \\
        \multicolumn{1}{c|}{} & SwinUNet~\cite{cao2022swin} & 77.65 $\pm$ 0.33 & 72.68 $\pm$ 0.42 & 81.58 $\pm$ 0.22 & 73.24 $\pm$ 0.27 & 86.99 $\pm$ 0.28 & 90.48 $\pm$ 0.14 \\
        \multicolumn{1}{c|}{} & MISSFormer~\cite{huang2022missformer} & 82.78 $\pm$ 0.39 & 77.01 $\pm$ 0.21 & 86.78 $\pm$ 0.13 & 83.92 $\pm$ 0.55 & 90.17 $\pm$ 0.66 & 91.73 $\pm$ 0.26 \\
        \multicolumn{1}{c|}{} & H2Former~\cite{he2023h2former} & 82.30 $\pm$ 10.69 & 76.10 $\pm$ 0.42 & 86.15 $\pm$ 0.28 & 82.59 $\pm$ 1.09 & 90.56 $\pm$ 0.85 & 91.07 $\pm$ 0.52 \\
       \multicolumn{1}{c|}{} & TransUnet~\cite{chen2021transunet} & 84.68 $\pm$ 0.40 & 78.44 $\pm$ 0.31 & 87.71 $\pm$ 0.20 & 85.23 $\pm$ 0.80 & 90.70 $\pm$ 0.57 & 92.38 $\pm$ 0.38\\ 
       \multicolumn{1}{c|}{} &  nnUNet~\cite{isensee2021nnu} & 87.54 $\pm$ 0.25 & 81.70 $\pm$ 0.23 & 89.69 $\pm$ 0.14 & 88.45 $\pm$ 0.28 & 91.17 $\pm$ 0.17 & 94.01 $\pm$ 0.14 \\ \midrule
       \multicolumn{1}{c|}{\multirow{9}{*}{\rotatebox{90}{Sequence + MinIP}}} & nnUNet\textsuperscript{*}~\cite{isensee2021nnu} & 88.25 $\pm$ 0.15 & 82.34 $\pm$ 0.17 & 90.07 $\pm$ 0.01 & 88.65 $\pm$ 0.32 & 91.72 $\pm$ 0.31 & 94.12 $\pm$ 0.15\\ 
       \multicolumn{1}{c|}{} & UNet\textsuperscript{*}~\cite{ronneberger2015u} & 78.56 $\pm$ 1.13 & 71.70 $\pm$ 1.24 & 83.17 $\pm$ 0.86 & 88.38 $\pm$ 1.73 & 79.76 $\pm$ 2.26 & 93.58 $\pm$ 0.79\\ 
       \multicolumn{1}{c|}{} & MISSFormer\textsuperscript{*}~\cite{huang2022missformer} & 80.95 $\pm$ 1.11 & 73.29 $\pm$ 1.34 & 84.34 $\pm$ 0.90 & 90.66 $\pm$ 0.64 & 79.54 $\pm$ 1.82 & 94.71 $\pm$ 0.27\\
       \multicolumn{1}{c|}{} & H2Former\textsuperscript{*}~\cite{he2023h2former} & 81.78 $\pm$ 0.39 & 74.05 $\pm$ 0.43 & 84.84 $\pm$ 0.30 & 87.89 $\pm$ 0.20 & 82.76 $\pm$ 0.43 & 93.46 $\pm$ 0.10\\
       \multicolumn{1}{c|}{} & DeepMedic~\cite{kamnitsas2017efficient} & 79.53 $\pm$ 0.89 & 73.51 $\pm$ 0.88 & 84.31 $\pm$ 0.64 & 85.20 $\pm$ 3.13 & 84.21 $\pm$ 3.43 & 92.20 $\pm$ 1.44\\
       \multicolumn{1}{c|}{} &
       DSNet~\cite{xie2024dsnet} & 87.41 $\pm$ 0.10 & 81.65 $\pm$ 0.11 & 89.65 $\pm$ 0.07 & 88.24 $\pm$ 0.09 & 91.34 $\pm$ 0.19 & 93.91 $\pm$ 0.04 \\
       \multicolumn{1}{c|}{} &
       CAVE~\cite{su2024cave} & 83.07 $\pm$ 1.88 & 75.89 $\pm$ 2.07 & 85.98 $\pm$ 1.44 & 87.05 $\pm$ 3.61 & 85.97 $\pm$ 1.64 & 93.15 $\pm$ 1.75 \\
       \multicolumn{1}{c|}{} &
       VSSNet~\cite{liu2024dias} & 82.72 $\pm$ 0.89 & 75.38 $\pm$ 0.90 & 85.67 $\pm$ 0.58 & \textbf{91.65} $\pm$ \textbf{0.72} & 81.37 $\pm$ 1.48 & \textbf{95.26} $\pm$ \textbf{0.31} \\
       \multicolumn{1}{c|}{} & DSANet & \textbf{88.58} $\pm$ \textbf{0.13} & \textbf{82.76} $\pm$ \textbf{0.16} & \textbf{90.33} $\pm$ \textbf{0.09} & 89.30 $\pm$ 0.23 & 91.56 $\pm$ 0.32 & 94.44 $\pm$ 0.11 \\ \bottomrule
    \end{tabular}
    \begin{tablenotes}
    \item '*' represents that the model input is a sequence and its MinIP.
    \end{tablenotes}
\end{table*}

\begin{figure*}[!t]
	\centering
	\includegraphics[width=\textwidth]{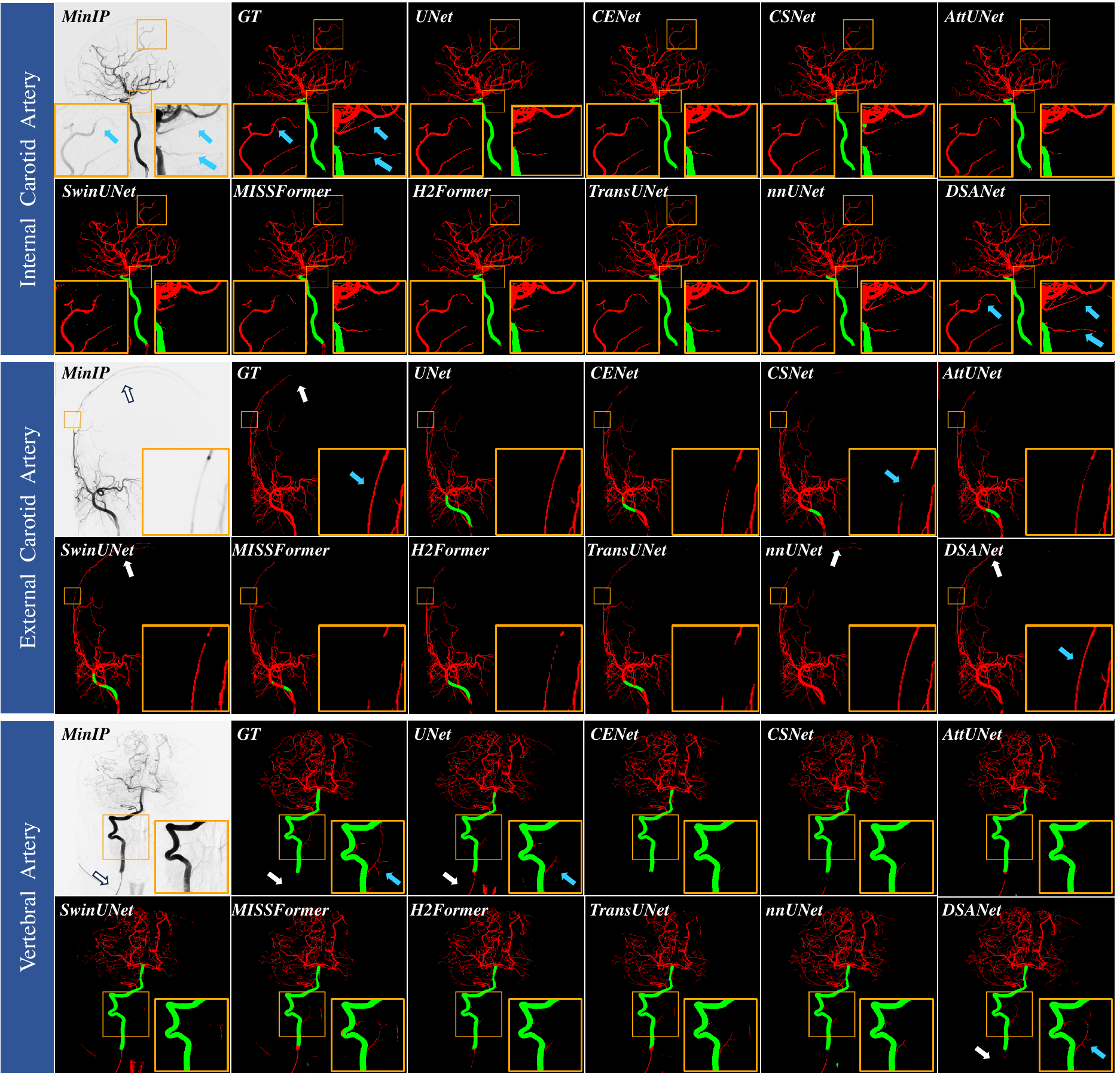}
	\caption{Segmentation results on the DSCA dataset. Green is MAT, and red is BV. Enlarged viewing for better clarity.}
	\label{fig-comparative}
\end{figure*}

\subsection{Cerebrovascular Segmentation Results}
We extensively assessed our proposed method on the DSCA dataset, comparing it with state-of-the-art segmentation approaches, which include UNet~\cite{ronneberger2015u}, CENet~\cite{gu2019net}, CSNet~\cite{mou2021cs2}, AttentionUNet~\cite{schlemper2019attention}, SwinUNet~\cite{cao2022swin}, MISSFormer~\cite{huang2022missformer}, H2Former~\cite{he2023h2former}, TransUNet~\cite{chen2021transunet}, nnUNet~\cite{isensee2021nnu}, ERNet~\cite{xu2023ernet}. 
Additionally, we also reproduced \cite{zhang2020neural} and \cite{meng2020multiscale} for comparison, as their codes are not publicly available. The comparative results of segmentation are shown in Table \ref{tab:comparative-experiment}.

We can observe that DSANet achieves the best performance across all metrics for segmenting both MAT and BV, which indicates the superiority of our method. 
In MAT segmentation, DSANet holds a significant advantage over other methods in terms of JAC, Dice, SEN, and AUC, surpassing H2Former by 4.32\%, 3.92\%, 4.50\%, and 3.39\%, respectively. A statistical significance of (\textit{p}$<$0.05) was obtained using a paired t-test between the two methods. 
In BV segmentation, the pixels of BV constitute a relatively small proportion of the total image pixels, which explains why our proposed method is not significantly ahead of the comparative method. 
Nevertheless, DSANet surpasses nnUNet by 1.33\% in JAC and 1.11\% in SEN with statistical significance (\textit{p}$<$0.05). Similarly, it also outperforms nnUNet by 0.87\% in Dice and 0.57\% in AUC, highlighting its superior performance with statistical significance (\textit{p}$<$0.05). 
 
For a more intuitive comparison of CA segmentation performance across different methods, we also provide a visual comparison in Fig. \ref{fig-comparative}. As we can see the shape and background of the CA are very intricate, with low contrast in small vessels. In this challenging context, our DSANet effectively segments the CA, particularly ensuring small vessels with more robust connectivity,
as indicated in the detailed orange boxes in ICA, ECA, and VA.
In addition, skulls with deep contrast are mistakenly extracted as vessels in other methods, 
while our method successfully suppresses such false positives, as shown by the white arrows in ECA and VA in Fig. \ref{fig-comparative}. 
In ICA, CENet and TransUNet perform well in segmenting vessels within the left orange box, while their performance diminishes in the right orange box. In contrast, DSANet achieves superior segmentation performance in both boxes. Furthermore, other methods erroneously segment the MAT in ECA (depicted in green), while DSANet effectively addresses this issue. 

We also compare the performance of the entire CA without distinguishing BV and MAT to demonstrate the superiority of DSANet further. As indicated in Table \ref{tab:comparative_all}, DSANet holds a slight advantage over other methods in JAC, Dice, SEN, and AUC, surpassing the second-place method by 1.06\%, 0.64\%, 0.85\%, and 0.43\%, respectively, with statistical significance (\textit{p}$<$0.05). In PRE, AttentionUNet outperforms our method, but it falls behind our method in SEN. This indicates that AttentionUNet may miss some positive samples, reducing sensitivity, whereas our method achieves a better balance between SEN and PRE. To achieve precise segmentation, it is essential to focus on the intricate details of small vessel structures, which represent a minority of the total image pixels. Thus, clDice~\cite{shit2021cldice} is employed to assess vessel connectivity. DSANet significantly outperforms other methods, surpassing nnUNet by 1.04\% in clDice with statistical significance (\textit{p}$<$0.001).

Furthermore, for a fairer comparison, we have added experimental comparisons with model inputs as a sequence and its corresponding MinIP image in Table~\ref{tab:comparative-experiment} and \ref{tab:comparative_all}. These comparisons include nnUNet~\cite{isensee2021nnu}, UNet~\cite{ronneberger2015u}, MISSFormer~\cite{huang2022missformer}, H2Former~\cite{he2023h2former}, DeepMedic~\cite{kamnitsas2017efficient}, VSSNet~\cite{liu2024dias}, CAVE~\cite{su2024cave} and our previous work, DSNet~\cite{xie2024dsnet}. Interestingly, for models like UNet, MISSFormer, and H2Former, their performance on metrics such as DICE was actually worse when using sequence input compared to just using the MinIP images. On the other hand, from the SEN metric, all of these methods saw a significant improvement, indicating that the introduction of the sequence increased the recall rate of the network, though this may also result in a decrease in precision. This could be because sequences contain more redundant information and background noise, making it crucial to handle these effectively. nnUNet can handle this issue effectively with its excellent training capabilities. nnUNet is a general model that outperforms some models specifically designed for DSA images, demonstrating its superior performance and generalization capabilities. However, as shown in Table~\ref{tab:comparative_all}, VSSNet surpasses nnUNet in SEN and AUC, despite being behind in other metrics. Additionally, compared to the two methods specifically designed for training with DSA sequences, CAVE and VSSNet, our DSANet significantly outperforms them in JAC, Dice, and PRE as shown in Table~\ref{tab:comparative-experiment} and \ref{tab:comparative_all}. Furthermore, it is noteworthy that our method performs slightly lower than VSSNet in terms of SEN and AUC metrics as shown in Table~\ref{tab:comparative_all}, which indicates that VSSNet tends to prioritize recall while neglecting precision. This result suggests that VSSNet is more inclined to capture true positives, while it may also yield a higher rate of false positives. Our method used nnUNet as the baseline and introduced TF and STF components, which can effectively learn spatio-temporal information, thereby achieving better performance.

\begin{table*}
    \centering
    \caption{Ablation study of these components of the proposed DSANet. The results are average$\pm$standard deviation obtained from the five-fold cross-validation experiment. All metrics are expressed as percentages (\%).}
    \label{tab:ablation study}
    \resizebox{\textwidth}{!}{
    \renewcommand{\arraystretch}{1.25}
    \begin{tabular}{ccc|cccccccccc}
    \toprule
         \multicolumn{3}{c}{Ablation Sets} \vline & \multicolumn{2}{c}{JAC} & \multicolumn{2}{c}{Dice} & \multicolumn{2}{c}{SEN} & \multicolumn{2}{c}{PRE} &  \multicolumn{2}{c}{AUC} \\ \midrule
         TEB & TF & STF & BV & MAT & BV & MAT & BV & MAT & BV & MAT & BV & MAT\\ \midrule 
         \multicolumn{3}{c}{MinIP} \vline & 76.76 $\pm$ 0.45 & 82.54 $\pm$ 2.42 & 86.45 $\pm$ 0.31 & 86.22 $\pm$ 2.50 & 85.29 $\pm$ 0.45 & 85.70 $\pm$ 2.50 & 88.06 $\pm$ 0.59 & 88.08 $\pm$ 2.20 & 92.41 $\pm$ 0.22 & 89.87 $\pm$ 2.51\\
         \multicolumn{3}{c}{Sequence} \vline & 76.77 $\pm$ 0.40 & 84.91 $\pm$ 2.23 & 86.45 $\pm$ 0.27 & 89.11 $\pm$ 2.31 & 85.12 $\pm$ 0.30 & 88.00 $\pm$ 2.04 & 88.29 $\pm$ 0.58 & 91.87 $\pm$ 2.51 & 92.33 $\pm$ 0.15 & 92.15 $\pm$ 2.31\\
         \multicolumn{3}{c}{Sequence + MinIP} \vline & 77.52 $\pm$ 0.22 & 85.90 $\pm$ 2.10 & 86.94 $\pm$ 0.17 & 89.88 $\pm$ 2.29 & 85.73 $\pm$ 0.32 & 88.76 $\pm$ 2.09 & 88.56 $\pm$ 0.31 & 91.93 $\pm$ 2.57 & 92.64 $\pm$ 0.15 & 93.00 $\pm$ 1.88\\ \midrule
         \checkmark &  &  & 77.40 $\pm$ 0.26 & 84.02 $\pm$ 4.82 & 86.87 $\pm$ 0.18 & 88.25 $\pm$ 4.87 & 86.12 $\pm$ 0.44 & 87.20 $\pm$ 5.12 & 87.99 $\pm$ 0.67 & 91.03 $\pm$ 3.75 & 92.83 $\pm$ 0.21 & 91.30 $\pm$ 4.67 \\ 
         \checkmark & \checkmark &  & 78.02 $\pm$ 0.22 & 86.00 $\pm$ 1.43 & 87.25 $\pm$ 1.43 & 89.91 $\pm$ 1.52 & 86.56 $\pm$ 0.10 & \textbf{91.86} $\pm$ \textbf{1.05} & 88.30 $\pm$ 0.22 & 92.81 $\pm$ 1.32 & 93.05 $\pm$ 0.05 & 93.18 $\pm$ \textbf{1.19} \\
         \checkmark &  & \checkmark & 77.83 $\pm$ 0.33 & 86.14 $\pm$ 2.21 & 87.14 $\pm$ 0.21 & 89.99 $\pm$ 2.29 & \textbf{86.84} $\pm$ \textbf{0.60} & 88.99 $\pm$ 2.18 & 87.78 $\pm$ 0.66 & 93.03 $\pm$ 2.63 & \textbf{93.18} $\pm$ \textbf{0.29} & 93.34 $\pm$ 2.00 \\
         \checkmark & \checkmark & \checkmark & \textbf{78.09} $\pm$ \textbf{0.20} & \textbf{88.22} $\pm$ \textbf{1.52} & \textbf{87.32} $\pm$ \textbf{0.13} & \textbf{92.26} $\pm$ \textbf{1.56} & 86.40 $\pm$ 0.39 & 91.37 $\pm$ 1.57 & \textbf{88.59} $\pm$ \textbf{0.53} & \textbf{94.01} $\pm$ \textbf{1.40} & 92.98 $\pm$ 0.18 & \textbf{94.98} $\pm$ \textbf{1.39} \\ \bottomrule
    \end{tabular} }
\end{table*}

\begin{table*}
    \centering
    \caption{Ablation study of these components of the proposed DSANet for the entire vessel. The results are average$\pm$standard deviation obtained from the five-fold cross-validation experiment. All metrics are expressed as percentages (\%).}
    \label{tab:ablation_all}
    \begin{tabular}{ccc|cccccc}
    \toprule
        \multicolumn{3}{c}{Ablation Sets} \vline & \multirow{2}{*}{clDice} & 
         \multirow{2}{*}{JAC} & \multirow{2}{*}{Dice} & \multirow{2}{*}{SEN} & \multirow{2}{*}{PRE} & \multirow{2}{*}{AUC} \\ \cmidrule{1-3}
         TEB & TF & STF & & & & & \\ \midrule
         \multicolumn{3}{c}{MinIP} \vline & 87.54 $\pm$ 0.25 & 81.70 $\pm$ 0.23 & 89.69 $\pm$ 0.14 & 88.45 $\pm$ 0.28 & 91.17 $\pm$ 0.17 & 94.01 $\pm$ 0.14 \\
         \multicolumn{3}{c}{Sequence} \vline & 87.78 $\pm$ 0.19 & 81.67 $\pm$ 0.23 & 89.65 $\pm$ 0.14 & 88.21 $\pm$ 0.16 & 91.40 $\pm$ 0.27 & 93.89 $\pm$ 0.08 \\
         \multicolumn{3}{c}{Sequence + MinIP} \vline & 88.25 $\pm$ 0.15 & 82.23 $\pm$ 0.16 & 90.07 $\pm$ 0.10 & 88.65 $\pm$ 0.32 & \textbf{91.72} $\pm$ \textbf{0.31} & 94.06 $\pm$ 0.15 \\ \midrule
         \checkmark &  &  & 88.29 $\pm$ 0.11 & 82.39 $\pm$ 0.11 & 90.11 $\pm$ 0.07 & 89.13 $\pm$ 0.37 & 91.30 $\pm$ 0.32 & 94.34 $\pm$ 0.17 \\ 
         \checkmark & \checkmark &  & 88.42 $\pm$ 0.13 & 82.60 $\pm$ 0.15 & 90.19 $\pm$ 0.09 & 89.28 $\pm$ 0.10 & 91.43 $\pm$ 0.25 & 94.40 $\pm$ 0.05 \\
         \checkmark &  & \checkmark & 88.41 $\pm$ 0.18 & 82.56 $\pm$ 0.24 & 90.17 $\pm$ 0.14 & \textbf{89.60} $\pm$ \textbf{0.34} & 91.12 $\pm$ 0.39 & 94.38 $\pm$ 0.16 \\
         \checkmark & \checkmark & \checkmark &
         \textbf{88.58} $\pm$ \textbf{0.13} & \textbf{82.76} $\pm$ \textbf{0.16} & \textbf{90.33} $\pm$ \textbf{0.09} & 89.30 $\pm$ 0.23 & 91.56 $\pm$ 0.32 & \textbf{94.44} $\pm$ \textbf{0.11} \\ \bottomrule
    \end{tabular}
\end{table*}

\subsection{Ablation Study}
We performed extensive ablation experiments on the DSCA dataset to illustrate the effectiveness of the components in DSANet. Utilizing nnUNet as the baseline, we systematically evaluated the influence of each component. 

\subsubsection{Ablation for MinIP and sequence} In Table~\ref{tab:ablation study} and \ref{tab:ablation_all}, the ablation experiments for MinIP and sequence are performed. As shown from the third to the fifth row of TABLE~\ref{tab:ablation study} and \ref{tab:ablation_all}, there are slight differences in performance between MinIP input and sequence input. While the sequence input does not bring additional spatial information, it introduces temporal information. MinIP, although not adding extra information, contains features that are easier to extract than the sequence input. Using both MinIP and sequence inputs together results in some performance improvement. Interestingly, for MAT, sequence input performs better than MinIP input, suggesting that the introduction of dynamic temporal information helps in distinguishing between MAT and BV more effectively.

\subsubsection{Ablation for temporal encoding branch} In this section, we present a comparative analysis of metrics including JAC, Dice, SEN, PRE, and AUC to assess the impact of integrating TEB. We incorporate TEB with nnUNet while maintaining consistent experimental settings. The ablation results, presented in the 4th row of Table~\ref{tab:ablation study}, demonstrate a significant improvement in both BV and MAT segmentation. 
This is primarily attributed to the integration of dynamic flow information. Notably, the incorporation of TEB resulted in an increase of over 1\% in JAC, Dice, SEN, PRE, and AUC in MAT. 

\subsubsection{Ablation for TemporalFormer} Subsequently, we validate the impact of TF based on the established foundation of the TEB. As depicted in the fifth row of Table~\ref{tab:ablation study}, all experimental outcomes have shown improvement to varying degrees compared to the fourth row. The enhancement in the segmentation results for MAT is more prominent than that for BV. Specifically, the SEN indicator has reached its optimal level in MAT in this method, reaching 91.86\%. 

\subsubsection{Ablation for spatio-temporal fusion module} 
Similarly, we assess the superiority of the STF module in conjunction with the TEB. Comparing the results in the sixth and fourth rows of Table \ref{tab:ablation study}, it is obvious that incorporating the STF module promotes better segmentation performance. Among them, the SEN and AUC metrics reach their optimal BV segmentation for our method, increasing the TEB by 0.72\% and 0.35\%, respectively. In the PRE metric, it slightly drops by 0.28\% and 0.21\% in BV, while outperforming them by 4.95\% and 2.00\% in MAT, compared to the baseline and the TEB, respectively. The last row in Table~\ref{tab:ablation study} outlines our results. A comprehensive review of the ablation experiments in the table explicitly illustrates the effectiveness of our method and its components. 
After adding both the TF and SFT components, our method falls behind the methods that only add one component in terms of the SEN metric. However, DSANet demonstrates superior performance in other metrics, such as the PRE metric. This suggests that adding one component increases the detection of false positives in the background. By combining these two components, our method is able to achieve better overall performance. As shown in Table \ref{tab:ablation_all}, these are the ablation study results for the entire vascular structure. After incorporating the sequence, both our method and nnUNet exhibited some performance improvement. However, after adding the TF and STF components, the performance difference was not very significant. We think this may be because the model takes into account the distinction between BV and MAT during training, leading to only a slight improvement in the overall vascular segmentation metrics. 

\begin{figure*}[!t]
	\centering
\includegraphics[width=\textwidth]{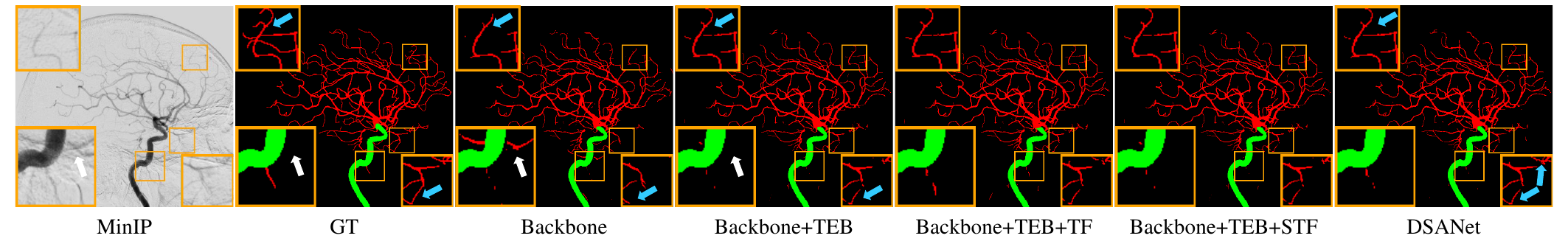}
	\caption{The ablation study results for baseline, DSANet, and the components of DSANet, with zoomed views for better visualization.}
	\label{fig-ablation}
\end{figure*}

In addition, we also provide visual comparisons in Fig.~\ref{fig-ablation}. The white arrows in the figure highlight a pseudo-shadow resembling a residual skull. While nnUNet misclassifies it as a vessel, our method successfully avoids such misclassification after the integration of TEB. Moreover, the small terminal vessels are better preserved in comparison with the baseline, as indicated by the blue arrows in the fourth column of Fig.~\ref{fig-ablation}. Additionally, through the integration of TF and STF in DSANet, the connectivity of small vessels shows obvious improvements, as depicted in the zoomed orange box and blue arrows of the last column of Fig.~\ref{fig-ablation}.

\begin{table*}[!t]
    \centering
    \caption{Experiments for TF in parallel, cldice loss, and variable length sequence. ALL represents the entire vessel.}
    \label{tab:ablation_others}
    \resizebox{\textwidth}{!}{
    \renewcommand{\arraystretch}{1.25}
    \begin{tabular}{c|ccccccccc}
    \toprule
         \multirow{2}{*}{Experiment Sets} & \multicolumn{3}{c}{JAC} & \multicolumn{3}{c}{Dice} & \multicolumn{3}{c}{AUC}  \\ \cmidrule{2-10}
         & BV & MAT & ALL & BV & MAT & ALL & BV & MAT & ALL\\ \midrule 
         Parallel & 77.89 $\pm$ 0.28 & 87.21 $\pm$ 1.90 & 82.78 $\pm$ 0.11 & 87.14 $\pm$ 0.22 & 90.93 $\pm$ 1.98 & \textbf{90.34} $\pm$ \textbf{0.07} & \textbf{92.99} $\pm$ \textbf{0.10} & 94.54 $\pm$ 1.49 & \textbf{94.45} $\pm$ \textbf{0.10} \\
         clDice & 66.04 $\pm$ 0.81 & 75.45 $\pm$ 1.34 & 74.65 $\pm$ 0.48 & 78.75 $\pm$ 0.58 & 81.32 $\pm$ 1.71 & 85.11 $\pm$ 0.32 & 87.80 $\pm$ 0.53 & 86.33 $\pm$ 1.20 & 90.68 $\pm$ 0.40 \\ 
         DSANet\_vl & 75.95 $\pm$ 0.30 & 78.28 $\pm$ 3.55 & 81.08 $\pm$ 0.32 & 85.78 $\pm$ 0.18 & 82.08 $\pm$ 3.73 & 89.25 $\pm$ 0.20 & 92.53 $\pm$ 0.30 & 86.33 $\pm$ 3.53 & 94.07 $\pm$ 0.16 \\ 
         DSANet\_fl & 75.05 $\pm$ 0.34 & 76.53 $\pm$ 3.68 & 80.53 $\pm$ 0.28 & 85.14 $\pm$ 0.25 & 80.44 $\pm$ 3.74 & 88.88 $\pm$ 0.19 & 92.02 $\pm$ 0.20 & 84.68 $\pm$ 3.68 & 93.70 $\pm$ 0.19 \\ \midrule
         DSANet & \textbf{78.09} $\pm$ \textbf{0.20} & \textbf{88.22} $\pm$ \textbf{1.52} & \textbf{82.76} $\pm$ \textbf{0.16} & \textbf{87.32} $\pm$ \textbf{0.13} & \textbf{92.26} $\pm$ \textbf{1.56} & 90.33 $\pm$ 0.23 & 92.98 $\pm$ 0.18 & \textbf{94.98} $\pm$ \textbf{1.39} & 94.44 $\pm$ 0.11  \\ \midrule
         CAVE\_vl & 69.12 $\pm$ 0.84 & 69.32 $\pm$ 2.25 & 76.50 $\pm$ 0.68 & 80.59 $\pm$ 0.89 & 74.46 $\pm$ 2.21 & 86.12 $\pm$ 0.60 & 88.68 $\pm$ 0.63 & 77.98 $\pm$ 2.03 & 91.44 $\pm$ 0.53 \\
         CAVE\_fl & 68.91 $\pm$ 3.00 & 73.11 $\pm$ 3.20 & 75.70 $\pm$ 2.30 & 80.87 $\pm$ 2.57 & 77.80 $\pm$ 3.21 & 85.78 $\pm$ 1.73 & 91.84 $\pm$ 2.74 & 80.83 $\pm$ 3.30 & 93.86 $\pm$ 2.17 \\ \midrule
         CAVE & 69.11 $\pm$ 2.80 & 79.09 $\pm$ 2.16 & 75.89 $\pm$ 2.07 & 81.12 $\pm$ 1.88 & 83.70 $\pm$ 2.15 & 85.98 $\pm$ 1.44 & 91.22 $\pm$ 2.21 & 86.84 $\pm$ 1.91 & 93.15 $\pm$ 1.75 \\
         \bottomrule
    \end{tabular} }
    \begin{tablenotes}
    \item 'vl' and 'fl' represent the variable length and fixed length, respectively.
    \end{tablenotes}
\end{table*}

\begin{table}
    \centering
    \caption{Experiments on DIAS dataset.}
    \label{tab:experiment_DIAS}
    \begin{tabular}{c|ccc}
    \toprule
         Methods & clDice & Dice & SEN\\ \midrule 
         nnUNet~\cite{isensee2021nnu} & 70.20 $\pm$ 4.65 & 78.09 $\pm$ 3.17 & 76.69 $\pm$ 7.81 \\
         VSSNet~\cite{liu2024dias} & 71.23 $\pm$ 5.65 & 78.34 $\pm$ 3.92 & 77.27 $\pm$ 7.73 \\  \midrule
         DSANet & \textbf{71.38} $\pm$ \textbf{5.27} & \textbf{78.37} $\pm$ \textbf{3.61} & \textbf{77.88} $\pm$ \textbf{8.15}  \\ \bottomrule
    \end{tabular}
\end{table}

\begin{figure}
	\centering
    \includegraphics[width=\linewidth]{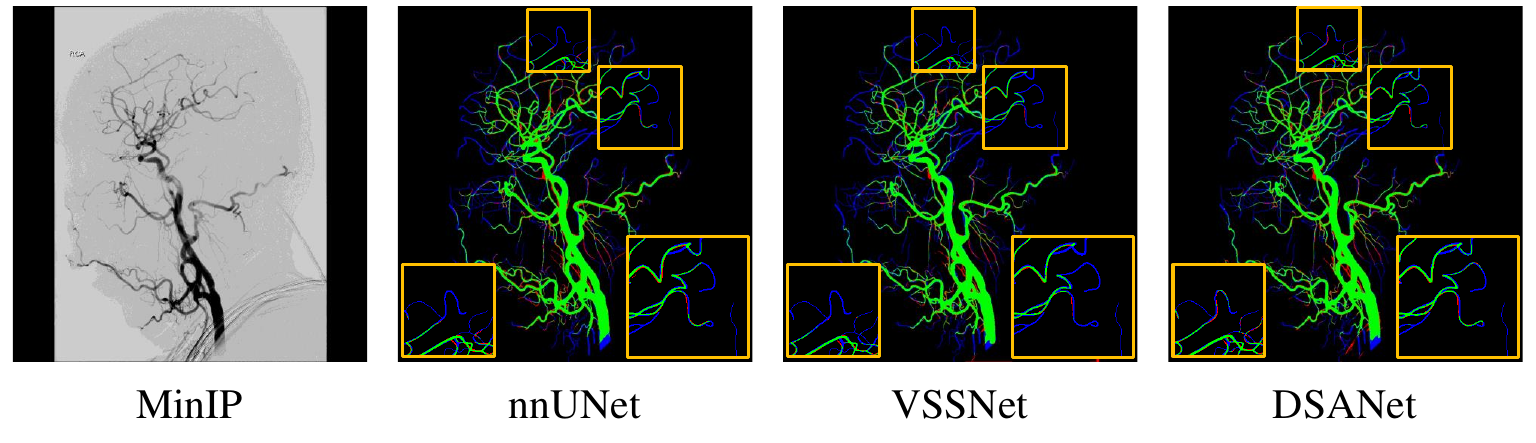}	
    \caption{Segmentation results on the DIAS dataset. Green, red, and blue represent true positive, false positive, and false negative, respectively.}
	\label{fig-dias}
\end{figure}

\section{Discussion}
\subsection{Additional Experimental Analysis}
\subsubsection{Experiment for TF in parallel} We compared the parallel execution of temporal attention and spatial attention in the TF component. The third row of Table~\ref{tab:ablation_others} demonstrates that performing in series is relatively better than parallel execution, especially in distinguishing between MAT and BV. We think directly combining the enhanced feature from the temporal and spatial dimensions could result in negative effects if performed in parallel.
\subsubsection{Experiment for clDice loss} 
clDice loss is a variation of Dice loss that encourages the network to focus more on the central topology of vessels. However, this emphasis can have unintended effects on larger vessels, potentially reducing segmentation accuracy for these structures. We have added an experiment in the fourth row of Table~\ref{tab:ablation_others} where we replaced CE loss and Dice loss with CE loss and clDice loss, but the results were not promising.
\subsubsection{Experiment for variable-length sequence input} While using variable-length sequence input may seem advantageous, vascular segmentation tasks typically use fixed-length inputs or a fixed number of channels. This approach is also employed in DIAS~\cite{liu2024dias}, where variable-length sequences are resampled to fixed-length sequences. Moreover, variable-length sequence inputs restrict the batch size to 1,  which is not conducive to model training. To verify the performance of raw sequence inputs, we have included an experiment with variable-length sequence input (DSANet\_vl), as shown in Table~\ref{tab:ablation_others}. Since nnUNet does not support variable-length sequences, we developed a new training framework. For a fair comparison, we also implemented fixed-length training (DSANet\_fl) by setting the batch size to 1. Similarly, we utilized CAVE to conduct the same training setting, resulting in both fixed-length (CAVE\_fl) and variable-length (CAVE\_vl) versions. Experimental results indicate that variable-length input outperforms fixed-length sequences, but is still behind the original DSANet.
\subsubsection{Experiment on the DIAS dataset} To further validate the generalization performance of our method, we conducted additional evaluations on the public DIAS dataset. To ensure a fair comparison, we retrained VSSNet on the DIAS dataset according to the setting in their paper. As presented in Table~\ref{tab:experiment_DIAS}, our DSANet demonstrates a slight performance improvement over both nnUNet and VSSNet. The qualitative results in Fig.~\ref{fig-dias} (highlighted in the orange box) illustrate that our method enhances cerebrovascular segmentation, underscoring its effectiveness and robustness in generalization.

\begin{figure}[!t]
	\centering
    \includegraphics[width=\linewidth]{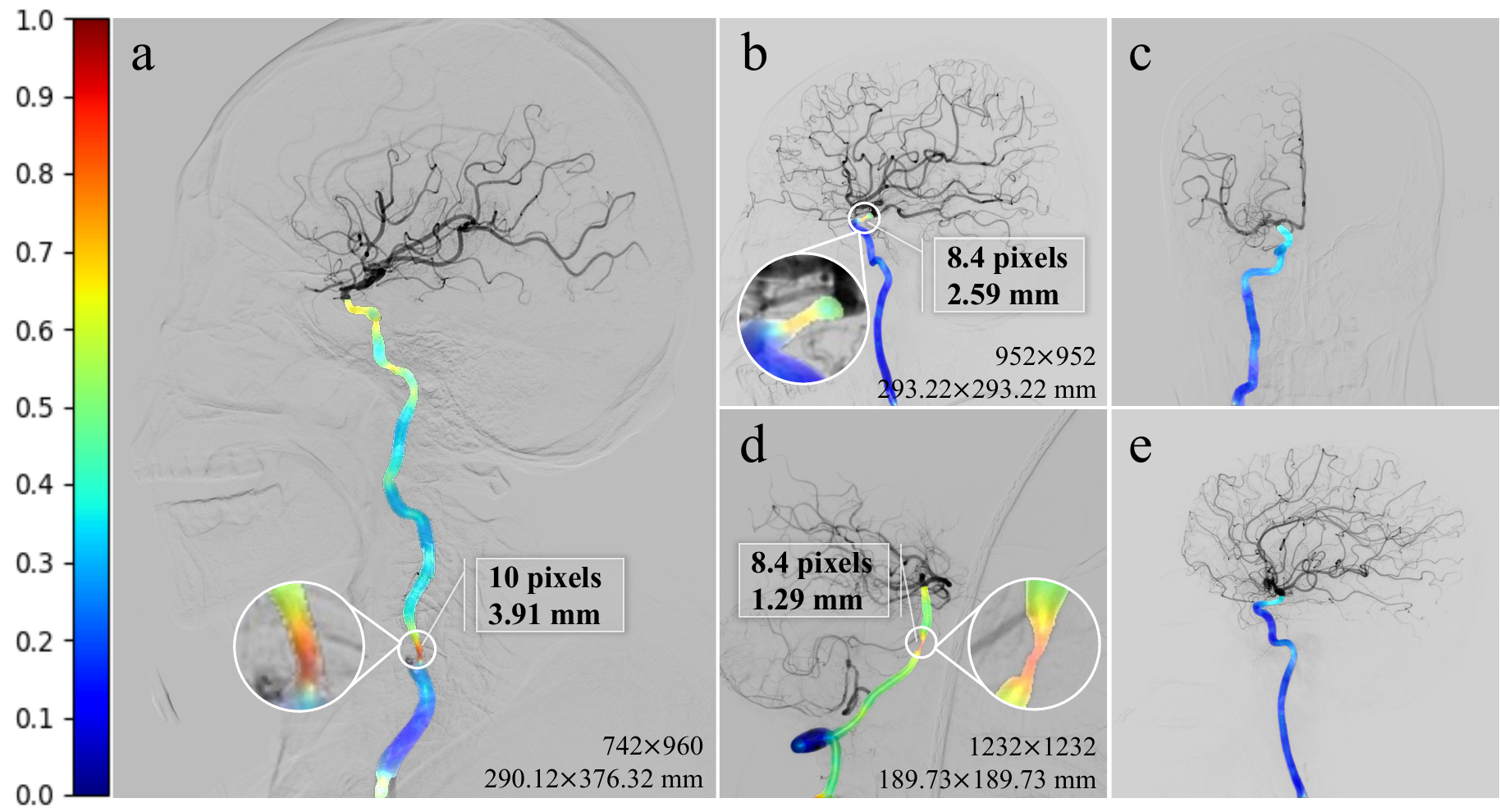}	
    \caption{Several instances where the MAT class is employed to assess stenosis. The more intense the red color, the more severe the stenosis.}
	\label{fig-stenosis}
\end{figure}

\subsection{DSCA: Promoting Clinical Cerebrovascular Analysis}
The DSCA dataset comprises 224 sequences, each meticulously annotated pixel by pixel using MinIP images. It includes ICA, ECA, and VA, with further subclassification of ICA and VA into BV and MAT. This classification closely aligns with the requirements of clinical analysis. Previous studies have demonstrated the value of precisely measuring carotid stenosis~\cite{rothwell1994prognostic}. The extraction of MAT facilitates automated diameter calculations for ICA and VA, particularly in evaluating biomarkers at stenosis sites. Fig.~\ref{fig-stenosis} reveals several instances where MAT is used to assess stenosis in the main trunk of the artery. In Fig.~\ref{fig-stenosis}-(a), (b), (d), the diameter of the stenosis region can be easily determined from the segmented MAT. Moreover, the color representation of the arterial main trunk appears normal in the absence of stenosis, as illustrated in Fig.~\ref{fig-stenosis}-(c), (e). 
Furthermore, it is well known that the ICA and VA are divided into segments~\cite{bouthillier1996segments,gailloud2022segmentation}. Some researchers have studied the tortuosity~\cite{koge2022internal,chen2020correlation}, morphometry~\cite{baz2021morphometry}, and diameter~\cite{benetos1985pulsed,bladin1995carotid} of the ICA and the clinical significance of specific ICA segments~\cite{cekic2024adventitia,kakizawa2000parameters}. Therefore, directly extracting the MAT is meaningful for these studies. Similarly, research on small vessels extends to studies on cerebral small vessel disease~\cite{keller2020large, pasi2020clinical} and cerebrovasculature labeling~\cite{chng2008territorial,robben2016simultaneous}, which remains a focus of ongoing clinical investigation. We believe our work can contribute to advancing these fields.

\subsection{Why Resample 8 Frames for Training?}
\begin{figure}[!t]
	\centering
\includegraphics[width=\linewidth]{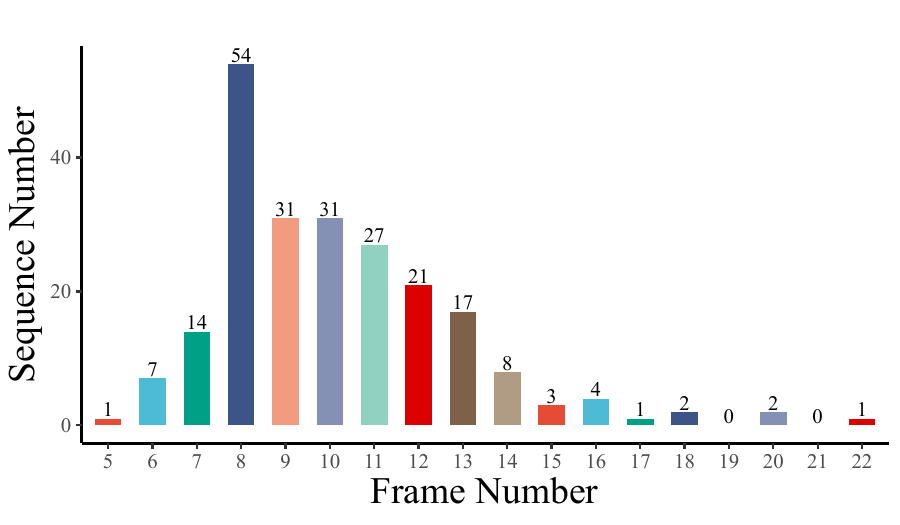}
	\caption{The distribution of the number of frames in DSCA.}
	\label{fig-data}
\end{figure}

\begin{figure}[!t]
	\centering
\includegraphics[width=\linewidth]{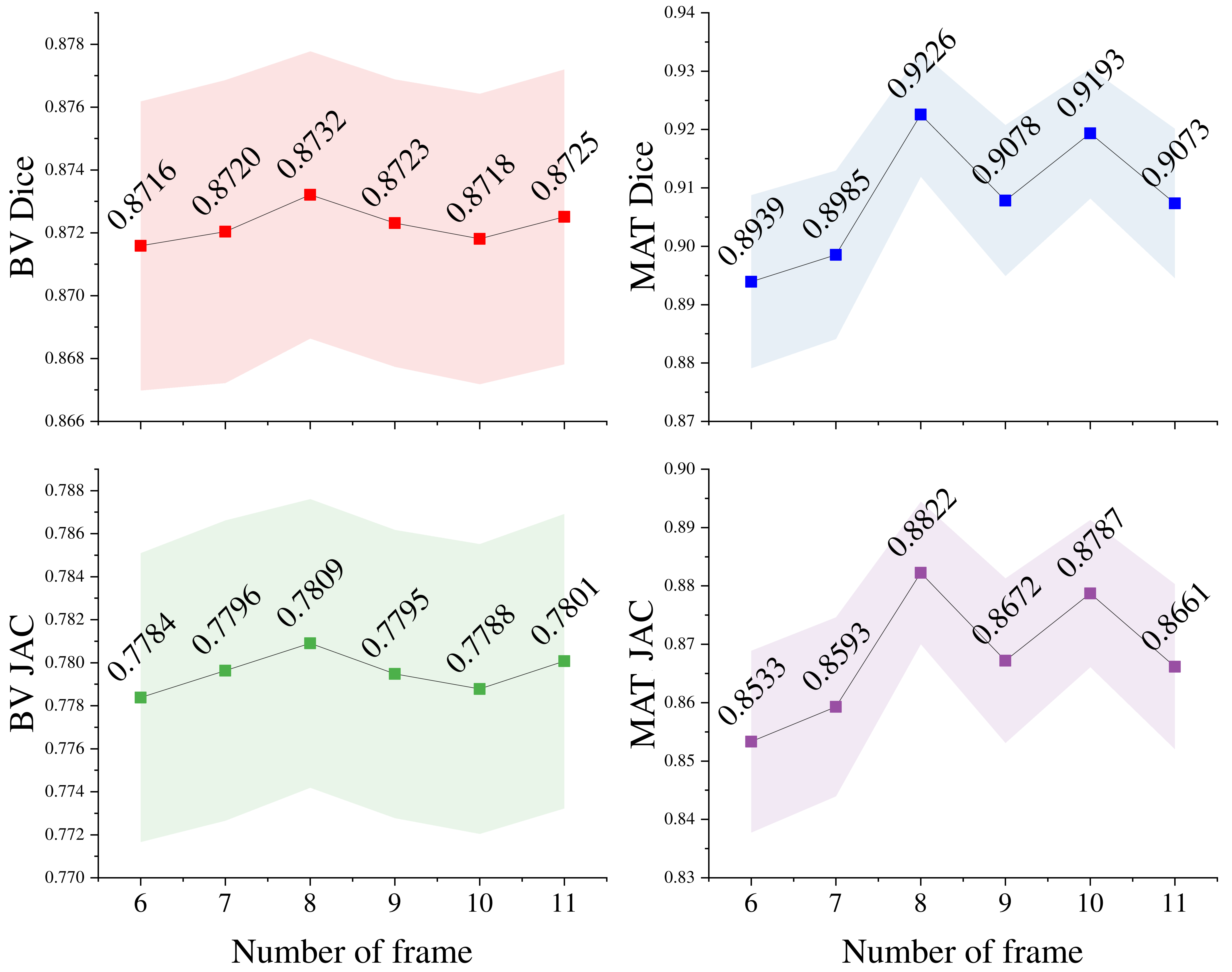}
	\caption{The influence of frame numbers on experiments. The evaluation metrics include Dice and JAC.}
	\label{fig-comparative_sample}
\end{figure}

To meet the input specifications of the model, we resampled the unfixed-length sequential frames to a fixed length. This procedure is necessary based on the observation that the majority of frames cluster around 8 frames, as illustrated in Fig. \ref{fig-data}. Resampling inputs that exceed 8 frames could introduce excessive resampling error, which in turn compromises the segmentation performance of the model. Conversely, resampling fewer than 8 frames results in a significant loss of temporal information in the DSA sequence, impeding the model's ability to capture sufficient dynamic flow features and similarly degrading segmentation performance. The experimental results, illustrated in Fig.~\ref{fig-comparative_sample}, confirm the finding that optimal segmentation performance is achieved when the input is resampled to 8 frames. Deviations from this resampling size, whether higher or lower, are associated with a decline in the model's segmentation efficacy. Consequently, we affirm that resampling consecutive frames to a length of 8 frames provides a comparatively optimal strategy.

\subsection{DSANet: Effectively Handle Complex Situations}
\begin{figure}[!t]
	\centering
	\includegraphics[width=0.9\linewidth]{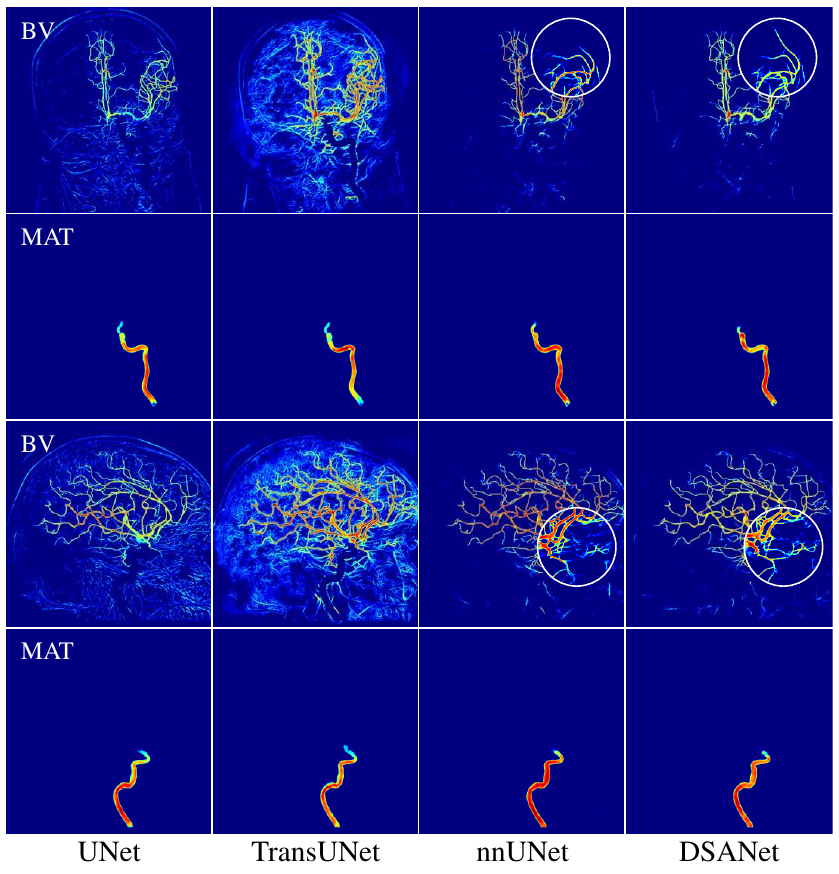}
	\caption{Heatmaps of the final layer output for different methods.}
	\label{fig-heatmap}
\end{figure}

It is challenging to accurately segment CA due to small vessels with low visibility and ambiguity between vessels and residual skull structures. To further validate the effectiveness of our approach, we generate heatmaps for the final layer output of the decoder and make comparisons with state-of-the-art methods, as shown in Fig.~\ref{fig-heatmap}. In the 2nd and 4th rows, all methods exhibit a positive response in MAT. Additionally, a comparison of the 1st and 3rd rows reveals that our DSANet produces a stronger response than UNet, TransUNet, and nnUNet on small vascular structures in BV, particularly in the areas highlighted by the white circles. Meanwhile, UNet and TransUNet are more susceptible to background artifacts and noise than the proposed DSANet, as shown in the 1st and 2nd columns of BV. This highlights the superior performance of our method in suppressing background interferences.

\section{Conclusion and Future Works}
In this paper, we have established the DSCA dataset, a large, carefully designed, and systematically annotated dataset for CA segmentation in DSA sequences. To our knowledge, the DSCA is the first publicly available DSA sequence dataset. It not only has pixel-wise vessel annotations but also includes subdivisions for BV and MAT, which are closely related to the clinical treatment of CVDs. This dataset will be highly beneficial for researchers in CVD studies. Furthermore, we have proposed DSANet, a spatio-temporal network for CA segmentation. Unlike 2D segmentation methods, DSANet incorporates an independent temporal encoding branch to capture dynamic temporal information. It integrates TemporalFormer and Spatio-Temporal Fusion modules to further process spatio-temporal features. Extensive experiments demonstrate the effectiveness of our method, particularly small vessels, holding significant potential for clinical applications. 

\bibliographystyle{IEEEbib}
\bibliography{refs}

\end{document}